\documentclass[aps,pra,twocolumn,showpacs,groupedaddress]{revtex4}

\usepackage{amsmath}
\usepackage{graphicx}
\usepackage{epsfig}
\usepackage{psfrag}

\usepackage{bm}

\usepackage{comment}

\usepackage[usenames]{color}

\newcommand{\be}{\begin{equation}}
\newcommand{\ee}{\end{equation}}
\def\beq#1\eeq{\begin{align}#1\end{align}}

\newcommand{\beeq}{\begin{eqnarray}}
\newcommand{\eeeq}{\end{eqnarray}}

\def\H1{\widehat{H}_1}

\newcommand{\lb}{\left[}
\newcommand{\rb}{\right]}
\newcommand{\lp}{\left(}
\newcommand{\rp}{\right)}

\def\op#1{{\Hat{\mathrm{#1}}}}

\def\bra#1{\ensuremath{\langle{#1}\vert}}
\def\ket#1{\ensuremath{\vert{#1}\rangle}}
\def\bracket#1#2{\ensuremath{%
    \langle{#1}\mkern1.2mu\vert\mkern1.2mu{#2}\rangle}}

\def\commutator#1#2{\mathinner{%
    \mathopen[#1,#2\mathclose]}}

\def\abs#1{\mathinner{\lvert#1\rvert}}

\begin{document}


\title{Exceptional and regular spectra of a generalized Rabi model}

\author{Michael Tomka$^{1}$, Omar El Araby$^{2}$, Mikhail Pletyukhov$^{3}$, Vladimir Gritsev$^{2}$}
\affiliation{
$^{1}$Department of Physics, Boston University, 590 Commonwealth Ave., Boston, MA 02215, USA\\
$^{2}$Institute for Theoretical Physics, University of Amsterdam, Science Park 904, Postbus 94485, 1098 XH Amsterdam, The Netherlands
$^{3}$Institute for Theory of Statistical Physics and JARA -- Fundamentals of Future Information Technology, RWTH Aachen, 52056 Aachen, Germany
}



\begin{abstract}
We study the spectrum of a generalized Rabi model in which co- and
counter-rotating terms have different coupling strengths.
It is also equivalent to the model of a two-dimensional electron gas in
a magnetic field with Rashba and Dresselhaus spin-orbit couplings. 
Like in case of the Rabi model, the spectrum of our generalized Rabi
model consists of the regular and the exceptional parts.
The latter is represented by the energy levels which cross at certain
parameter values which we determine explicitly.
The wave functions of these exceptional states are given by finite
order polynomials in the Bargmann representation. 
The roots of these polynomials satisfy a Bethe ansatz equation of the
Gaudin type.
At the exceptional points the model is therefore quasi-exactly
solvable.
An analytical approximation is derived for the regular part of the
spectrum in the weak- and strong-coupling limits. 
In particular, in the strong-coupling limit the spectrum consists of
two ladders of equidistant levels.
\end{abstract}


\maketitle

\section{Introduction}

The Rabi model~\cite{Rabi} is a fundamental model of light-matter
interaction.
It describes a single-mode photonic field interacting with a single
two-level emitter,
\be
  \op{H}_{\mathrm{R}} 
  = 
  \omega \op{a}^{\dag}\op{a}
  +
  \omega_{0} \op{\sigma}_{z}
  +
  g \lp \op{a} + \op{a}^{\dag} \rp \lp \op{\sigma}_{+} + \op{\sigma}_{-} \rp,
\label{eq:Rabi}
\ee   
where the bosonic operators $\op{a}, \op{a}^{\dag}$ describe the
photons, and $\op{\sigma}_{\mu}$, $\mu=z,\pm$, are the Pauli matrices
describing a two-level emitter.
When the coupling strength $g/\omega$ is small $\sim 10^{-2}$ and the
near-resonance condition is satisfied, $\omega \sim 2\omega_{0}$, 
it is legitimate to make the rotating wave approximation (RWA) by
neglecting the counter-rotating terms $\op{a} \, \op{\sigma}_{-}$ and
$\op{a}^{\dag} \op{\sigma}_{+}$.
In this case, known as the Jaynes-Cummings (JC) model~\cite{JC}, 
the operator of the total number of excitations 
$\op{N}_{\mathrm{ex}} = \op{a}^{\dag}\op{a} + \op{\sigma}_{+}\op{\sigma}_{-}$ 
is a conserved quantity which ensures exact solvability of the JC
model.
On the other hand, in the Rabi model the only conserved
quantity is the parity
$\op{\Pi} = \exp(i\pi\op{N}_{\mathrm{ex}})$.
The question of exact solvability of the Rabi model has been debated for a
long time, and the recent renewal of interest to the subject~\cite{Braak},~\cite{Morozov}
has been motivated by the rapid experimental progress in quantum optics.
Several regimes of the Rabi model~(\ref{eq:Rabi}) are usually
distinguished in the literature depending on the coupling strength or
the detuning $\Delta=\omega-2\omega_{0}$.
In terms of the dimensionless parameter $\eta = g/\omega$ these are: 
(i) the weak-coupling regime, when the JC model is applicable,
$\eta \sim 10^{-2}$; 
(ii) the strong-coupling regime, $10^{-2}<\eta<0.1$; 
(iii) the ultra-strong coupling regime, $0.1<\eta<1$, and 
(iv) deep strong-coupling regime $\eta>1$.
For sufficiently large detuning so that $|\Delta| \gg 2\omega_{0}$, 
the RWA breaks down even for a relatively weak coupling. 
If couplings of the field and the emitter to dissipative
baths ($\gamma$ and $\Gamma$, respectively) are included, it is often
assumed that the cooperativity factor $\xi = g^{2}/\gamma\Gamma$ is
large enough to ensure almost coherent short time evolution.
It is worth noting that the standard weak-coupling master (Lindblad) equation
approach to dissipative dynamics in the strong-coupling regime should
be taken with caution~\cite{Beaudoin}.
Namely, the reduced density matrix equation should be expressed in
terms of exact eigenstates of the isolated subsystem. 
This calls for detailed studies of the spectrum in the different limits (i)-(iv).
Experimentally, the weak-coupling regime is achieved in cavity QED 
setups~\cite{cav-QED}, while the regimes up to the ultra-strong
coupling have been recently accessed using circuit QED
systems~\cite{circ-QED-1},~\cite{circ-QED-2}.

The analytical solution of the Rabi model in terms of transcendental
functions has been found recently in Ref.~\cite{Braak}. On the other
hand, several analytical approximations are also available.
Thus, uniformly approximate results for energy levels valid in the
whole range of parameters were found in~\cite{FKU}; also known are the
approximation based on the polaron-like transformation, 
which is valid in the intermediate coupling (Bloch-Siegert)
regime~\cite{BS-regime}, the adiabatic approximation valid in the
strong-coupling regime~\cite{IGMS}, and the deep strong-coupling
approximation~\cite{casanova}.

A complementary information on the spectrum of the Rabi model is
provided by the {\it quasi-exact solutions} (QES). 
Indeed, it was
observed~\cite{Judd},~\cite{Reik},~\cite{Kus},~\cite{KL},~\cite{RD},~\cite{KKT},~\cite{EmaryBishop}
that the spectrum of the Rabi model has both regular and exceptional
pieces.
The exceptional parts of the spectrum are those whose wave functions are
finite-order polynomials in the Bargmann representation.
The energies of the exceptional solution are integer-valued
$E=n\omega-g^{2}/\omega$, where for every $n$ there is a special (polynomial)
condition on the model parameters for
which this solution is valid.
It was proven in~\cite{Kus} that two neighboring levels
cross on parallel straight lines $E=n\omega-g^{2}/\omega$ in
parameter space. 
Moreover, for each $n$ the number of such crossings is precisely $n$,
and there are no other crossings away from these lines.
The connection of exceptional solutions with the concept of quasi-exact
solvability (see~\cite{Turbiner} for an extensive review and references)
has been discussed in~\cite{KKT}.
We also note that in the quasi-classical regime the model exhibits
chaotic behavior, and the exceptional solutions correspond to the isolated set
of periodic orbits~\cite{chaos}.

In this paper we study a generalized Rabi model
\begin{align}
 \op{H}_{\mathrm{gR}}
 =&
 \omega \op{a}^{\dag}\op{a} 
 +
 \omega_{0} \op{\sigma}_{z}
 +
 g_{1} \lp \op{a}^{\dag}\op{\sigma}_{-} + \op{a}\op{\sigma}_{+} \rp \nonumber \\
 &+
 g_{2} \lp \op{a}^{\dag}\op{\sigma}_{+} + \op{a}\op{\sigma}_{-} \rp,
 \label{gR}
\end{align}
which interpolates between the JC model ($g_{2}=0$) and the
original Rabi model ($g_{1}=g_{2}$). 
There are several motivations to consider this model.
First, as observed in~\cite{EES} it can be mapped onto the model describing
a two-dimensional electron gas with Rashba ($\alpha_{R}\sim g_{1}$) 
and Dresselhaus ($\alpha_{D}\sim g_{2}$) spin-orbit couplings subject
to a perpendicular magnetic field (the Zeeman splitting thereby equals $2\omega_{0}$).
The Rashba spin-orbit coupling can be tuned by an applied electric
field while the Zeeman term is tuned by an applied magnetic field. 
This allows us to explore the whole parameter space of
the model.
Second, the model can directly emerge in quantum optics in the context of
cavity QED~\cite{SBOT} beyond the dipole approximation.
For example in Ref.~\cite{GriPar2013} a realization of the
generalized Rabi model~(\ref{gR}) based on resonant Raman transitions
in an atom interacting with a high finesse optical cavity mode is
proposed.

Here we describe the exceptional solutions of the model~(\ref{gR}). 
As in the Rabi model we find exceptional points corresponding to
doubly-degenerate level crossings in
parameter space ($\omega,\omega_{0},g_{1},g_{2}$).
These degeneracies (intersection points) form curves whose equations can be
determined from a set of algebraic conditions.
The level intersections occur only at integer values of the energy
$\frac{E}{\omega}+\frac{g_{1}^{2}+g_{2}^{2}}{2\omega^{2}}$, and
no intersections are observed elsewhere.
We discuss several interesting links between the structure of the
exceptional solutions and quasi-exact solvability, and the Gaudin-type Bethe ansatz solvable
models.
Namely, the conditions that the parameters of the generalized Rabi model need to
satisfy such that the energy levels are doubly-degenerate, are given in terms of Bethe ansatz
equations that have the same form as those of a reduced
Richardson model from superconductivity theory. The pairing
interaction strength of the conduction electrons corresponds then to
$\omega^{2}/(2g_{1}g_{2})$ from our generalized Rabi model.
Moreover, we analyze the weak- and strong-coupling limits of the
regular spectrum. 
In particular, we show that in the strong-coupling limit the spectrum
consists of two ladders of quasi-degenerate equidistant levels for
a small splitting of the two-level system, $\omega_{0}\ll\omega$.
Whereas for $\omega_{0}\gg\omega$ the spectrum is similar to the one
of the JC model.
We supplement our analytical study by comprehensive numerical
calculations.

The paper is organized as follows.
In Sec.~\ref{sec:exceptionalsolutionsforthegRabi} we present the 
procedure of determining the exceptional part of the spectrum of the
generalized Rabi model~(\ref{gR}).
It is shown that, as in the case of the Rabi model, the exceptional
part corresponds to doubly-degenerate level crossings
for which the associated eigenfunctions in Bargmann space are
polynomials of finite order.
Further, we establish explicitly the conditions on the system parameters
at which these level crossings occur.
We consider also the limits where the exceptional solutions can be
determined analytically.
In Sec.~\ref{sec:regularspectrum} we discuss two limiting cases of the
regular part of the spectrum, namely a weak coupling limit of
either small $g_{1}$ or small $g_{2}$ and a strong coupling limit
for large values of both $g_{1}$ and $g_{2}$.
Section~\ref{sec:conclusionanddiscussions} contains the conclusions of
the present work.

\section{Exceptional solutions for the generalized Rabi Hamiltonian}
\label{sec:exceptionalsolutionsforthegRabi}

\subsection{Hamiltonian in Bargmann representation}

To determine the exceptional solutions of the generalized
Rabi model~(\ref{gR}) we use the Bargmann representation for the
bosonic creation and annihilation operators in the space of analytic
functions in a complex variable $z$,
\be
    \op{a} \to \frac{d}{dz}, \qquad \op{a}^{\dagger} \to z.
\ee
Then, after applying the transformation $\op{\mathcal{H}}=
\op{P} \op{H}_{\mathrm{gR}}  \op{P}^{-1}$, with
\be
   \op{P}
   =
   \begin{pmatrix}
    -\frac{1}{2} \frac{\sqrt{g_{2}}}{\sqrt{g_{1}}} & \frac{1}{2} \\
    \frac{1}{2} \frac{\sqrt{g_{2}}}{\sqrt{g_{1}}} & \frac{1}{2}
   \end{pmatrix},
   \qquad
   \op{P}^{-1}
   =
   \begin{pmatrix}
    -\frac{\sqrt{g_{1}}}{\sqrt{g_{2}}} & \frac{\sqrt{g_{1}}}{\sqrt{g_{2}}} \\
    1 & 1
   \end{pmatrix},
\ee
the stationary Schr\"odinger equation $\op{\mathcal{H}} \psi = E \psi$
for the two component wave function,
\be
   \psi(z)
   =
   \begin{pmatrix}
    \psi_{1}(z) \\
    \psi_{2}(z)
   \end{pmatrix},
\ee
becomes a system of two first-order linear differential equations
for the functions $\psi_{1}(z)$ and $\psi_{2}(z)$
\begin{align}
  \label{eq:psizonep}
  & \lp z - \nu \rp \frac{d\psi_{1}}{dz}
  - \lp \frac{\lambda_{+}}{\nu} z + e \rp \psi_{1}
  + \lp \frac{\lambda_{-}}{\nu} z - \delta \rp \psi_{2}
  =0,
  \\
  \label{eq:psiztwop}
  & \lp z + \nu \rp \frac{d\psi_{2}}{dz}
  + \lp \frac{\lambda_{+}}{\nu} z - e \rp \psi_{2}
  - \lp \frac{\lambda_{-}}{\nu} z + \delta \rp \psi_{1}
  = 0,
\end{align}
where we introduced the dimensionless quantities
\begin{align}
    \delta &\equiv \frac{\omega_{0}}{\omega}, \quad
    \lambda_{\pm} \equiv \frac{\frac{1}{2}\lp g_{1}^{2} \pm g_{2}^{2} \rp}{\omega^{2}}, \quad
    \nu \equiv \frac{\sqrt{g_{1}g_{2}}}{\omega},  \nonumber \\
    e &\equiv \frac{E}{\omega}, \quad
    \epsilon \equiv e + \lambda_{+}.
\end{align}

In analogy to the Rabi model~\cite{Kus} we study the analytical
properties of the solutions $\psi_{1}(z)$ and $\psi_{2}(z)$
around the two singular points $z=\pm\nu$.
To this end we expand the solutions as power series about one of the
singular points $z=\nu$:
\be
  \psi_{i}(z) = (z-\nu)^{s} \sum_{n=0}^{\infty} c_{n}^{(i)}
  (z-\nu)^{n}, \qquad i=1,2.
  \label{eq:taylor}
\ee
Inserting this expansion into Eqs.~(\ref{eq:psizonep})
and~(\ref{eq:psiztwop}) yields the so-called indicial equation
\be
  s(\lambda_{+}+e-s)=0,
  \label{eq:indicialequation}
\ee
for the possible values of $s$.
We note that the same condition is found for the singularity
$z=-\nu$.
The first solution $s=0$ of Eq.~(\ref{eq:indicialequation}) shows that 
we can always find an analytic solution
$\psi(z)=(\psi_{1}(z),\psi_{2}(z))^{T}$ in a neighborhood of the
singularities $z=\pm\nu$.
The second solution $s=\lambda_{+}+e$ implies that another linearly
independent analytic solution $\tilde{\psi}(z)$ can occur and then the
energy level is doubly degenerate. But this second solution is only
analytic if the energy satisfies $e=n-\lambda_{+}$, where $n$ is a non-negative
integer, and in this case is given by
$\tilde{\psi}(z)=(\psi_{2}(-z), \psi_{1}(-z))^{T}$.
The exact condition on the other parameters for which these doubly degenerated
exceptional solutions appear will be determined in the following. 
We will show that these solutions are polynomials of
finite order.

Differentiating Eq.~(\ref{eq:psizonep}) one more time and
eliminating $\psi_{2}(z)$ from Eq.~(\ref{eq:psiztwop}) and $\psi_{2}'(z)$
from Eq.~(\ref{eq:psizonep}),
we get a second-order differential equation for $\psi_{1}(z)$.
After some transformations (see Appendix~\ref{appendix:A}) these
equations can be written as
\be
 \lb 
  \frac{d^{2}}{dz^{2}}
  + 
  \lp
   \sum_{s=1}^{3}\frac{\nu_{s}}{z-\rho_{s}} + \nu_{0} 
  \rp \frac{d}{dz}
  + 
  \frac{D_{2}(z)}{\prod_{s=1}^{3}(z-\rho_{s})} 
 \rb \chi(z) = 0,
 \label{eq:diffeqchisimp}
\ee
where $D_{2}(z)=\sum_{s=0}^{2}d_{s}z^{s}$ is a polynomial of degree 2
with coefficients given by 
\begin{align}
  d_{0}
  &=
  \kappa
  \lp
     \delta^{2}
   - \epsilon^{2}
   + 2 \epsilon \lambda_{+}
   - \lambda_{+}^{2}
   + \lambda_{+}
   + \nu^{2}
   + \nu^{4}
  \rp
  \nonumber \\
  &+ \nu \lp \epsilon-\lambda_{+}-\nu^{2}\rp,
  \\
  d_{1}
  &=
  e(e+1)-\delta^{2}
  + 
  \delta \frac{\lambda_{+}}{\lambda_{-}}
  +
  \nu \kappa
  -
  \nu^{2}
  -
  2 \nu \epsilon \kappa
  -
  \nu^{4},\\
  \label{eq:defd2}
  d_{2} &= 2 \nu \epsilon,
\end{align}
and $\chi(z)=\exp(\nu z)\psi_{1}(z)$.
The other constants in Eq.~(\ref{eq:diffeqchisimp}) are
\begin{align}
  \rho_{1} &= \nu,
  \qquad
  \rho_{2} = -\nu,
  \qquad
  \rho_{3} = \kappa,
  \\
  \nu_{1} &= -\epsilon + 1,
  \quad
  \nu_{2} = -\epsilon,
  \quad
  \nu_{3} = -1,
  \quad
  \nu_{0} = -2 \nu,
\end{align}
where we set
\be
  \kappa
  \equiv
  \frac{\delta \nu}{\lambda_{-}}
  =
  \frac{2 \omega_{0} \sqrt{g_{1}g_{2}}}{g_{1}^{2} - g_{2}^{2}}.
\ee
Note that the differential equation~(\ref{eq:diffeqchisimp}) 
is more general than the one corresponding to the usual Rabi model
with $g_{1}=g_{2}=g$,
yet it also has a polynomial solution
\be
 \chi(z)=\prod_{i=1}^{n}(z-z_{i})
 \label{eq:polyan}
\ee 
of degree $n$, if the coefficients $d_{j}$ satisfy certain relations.
These were explicitly found~\cite{zhang} 
using the functional Bethe ansatz method~\cite{sklyanin}.
This method simply consists in inserting
$\chi(z)$ into Eq.~(\ref{eq:diffeqchisimp}) and then dividing by
$\chi(z)$. 
The resulting equation gives then rise to the conditions that the
coefficients $d_{j}$ need to satisfy such that $\chi(z)$ is a valid
solution of Eq.~(\ref{eq:diffeqchisimp}).
In our case these conditions read
\begin{widetext}
\begin{align}
 d_{2} &= 2 \nu n, \label{1cond} \\
 d_{1} &= 2 \nu Z_{1} - n \lb (n-1) + \sum_{s=1}^{3}\nu_{s} + 2\nu \sum_{s=1}^{3}\rho_{s} \rb, \label{2cond}\\
 d_{0}
 &=
 2 \nu Z_{2} - \lb 2(n-1) + \sum_{s=1}^{3}\nu_{s} + 2\nu \sum_{s=1}^{3}\rho_{s} \rb Z_{1}
 +
 n(n-1) \sum_{s=1}^{3} \rho_{s}
 +
 n \lb 2\nu \sum_{s<p}^{3} \rho_{s}\rho_{p} + \sum_{s \neq p \neq
   q}^{3} \nu_{s}(\rho_{p}+\rho_{q}) \rb, \label{3cond}
\end{align}
\end{widetext}
where $Z_{k} = \sum_{i=1}^{n}z_{i}^{k}$ and $z_{i}$ are the roots of the
Bethe ansatz equations
\be
  \sum_{j \neq i}^{n} \frac{2}{z_{i}-z_{j}}
  +
  \sum_{s=1}^{3}\frac{\nu_{s}}{z_{i}-\rho_{s}}
  +
  \nu_{0}
  =
  0,
\ee
explicitly
\be
  \sum_{j \neq i}^{n} \frac{2}{z_{j}-z_{i}}
  +
  \frac{\epsilon-1}{z_{i}-\nu}
  +
  \frac{\epsilon}{z_{i}+\nu}
  +
  \frac{1}{z_{i}-\kappa}
  +
  2\nu
  =
  0,
  \label{BAeq}
\ee
with $i= 1, 2, \ldots, n$.
Eqs.~(\ref{eq:defd2}) and~(\ref{1cond}) yield the allowed energy
spectrum,
\be
  \epsilon = n, \quad \mbox{or} \quad E = \omega (n-\lambda_{+}). 
  \label{eq:energy-except}
\ee
Substituting this into the second and the third conditions,
Eqs.~(\ref{2cond}),~(\ref{3cond}), gives
\begin{align}
  2 \nu Z_{1}
  &=
  \lambda_{+}^{2}
  -
  \lp 2n+1-\frac{\kappa}{\nu} \rp \lambda_{+}
  -
  \lp \delta^{2} + \nu (\nu-\kappa)+\nu^{4} \rp, 
  \label{txfz1-z2} \\
  2 \nu^{2} Z_{2}
  &=
  - \lambda_{+}^{2}
  + (2n + 1 - \frac{\kappa}{\nu} + \kappa^{2} - \nu^{2}) \lambda_{+}
  \nonumber \\
  &+ (\delta^{2} + \nu (\nu  - \kappa) + \kappa^{2}\nu^{2} + 2n \nu^{2}(\nu^{2}+1)),
 \label{txsz1-z2}
\end{align}
where
$\lambda_{+}=\sqrt{\delta^{2}\frac{\nu^{2}}{\kappa^{2}}+\nu^{4}}$,
which comes from the identity $\lambda_{+}^{2}-\lambda_{-}^{2}=\nu^{4}$.
A derivation of those formulas can be found in
Appendix~\ref{appendix:B}.


\subsection{Analysis of the spectrum: exceptional case}

The procedure for determining the locations of the exceptional
solutions in the parameter space is now the following:
by fixing the number of nodes $n$ of the eigenfunctions $\chi(z)$ and
three out of the four parameters $(\omega,\omega_{0},g_{1},g_{2})$, we
solve the Bethe ansatz equations~(\ref{BAeq}) according to the method
proposed in~\cite{FAG} under the
conditions~(\ref{eq:energy-except}),~(\ref{txfz1-z2})
and~(\ref{txsz1-z2}).
This yields a polynomial equation for the remaining
parameter.
The solutions of this polynomial equation provides us the
values of the remaining parameter for which the eigenfunctions $\chi(z)$ are
given by a polynomial of order $n$ in the Bargmann representation.
A detailed explanation of this procedure is presented in
Appendix~\ref{appendix:C}.
The condition for an existence of polynomial solutions implies that
the two solutions $\psi(z)$ and $\tilde{\psi}(z)$ mentioned after
Eq.~(\ref{eq:indicialequation})
are degenerate with the eigenenergies given
by~(\ref{eq:energy-except}).
Away from these exceptional points these degeneracies are lifted.
The exceptional solutions correspond therefore to the
doubly-degenerate energy level crossings in parameter space.
We note that the polynomial solutions obtained for the exceptional
part of the spectrum can be related to the generalized Heine-Stieltjes
polynomials~\cite{Links}.

It is interesting to realize that the Bethe ansatz
equations~(\ref{BAeq}) have the same form as those for the reduced BCS
(Richardson) model having three degenerate levels of energies
$\rho_{1,2,3}$ with degeneracies $\nu_{1,2,3}$ respectively. 
This corresponding physical model is integrable and can be derived
from the generalized Gaudin models (see, e.g.,~\cite{genGaud} for
review).
Interestingly, the energy of that reduced BCS model is proportional to
$Z_{1}$ up to an additive constant. 
We would like to point out that there is no known mapping between the two models. We therefore understand this connection rather as a generic mathematical structure behind Gaudin-type models and polynomial solutions of the differential equations. This common structure is nothing else than the electrostatic analogy which has been discussed extensively in the literature, see e.g.~\cite{Ismail} for the case of differential equations and~\cite{genGaud} for Gaudin-type models.%

In general, the Bethe equations can be analyzed using the mapping to
the Riccati hierarchy~\cite{FAG}. The case of $\kappa^{2}=\nu^{2}$
requires special attention. In this case $\delta=\pm \lambda_{-}$ the
Bethe ansatz equations are those of the degenerate two-step model~\cite{ML}. Namely, when $\delta=-\lambda_{-}$ (so that $\kappa=-\nu$) the three roots $\rho_{1,2,3}$ degenerate into two (namely to $\pm\nu$) and moreover, the polynomial $D_{2}(z)$ is factorized as $D_{2}(z)=(z+\nu)(2\nu\epsilon z+d_{0}/\nu)$ which simplifies the differential and the Bethe ansatz equations. The corresponding conditions are given in Ref.~\cite{zhang}, Eqs.~A.12-A.14 for $\sigma=0$.

One of the central results of this paper is that the conditions
determining the locations of the exceptional solutions in parameter
space are given through the Bethe ansatz equations~(\ref{BAeq}), which
are the same as those of the reduced BCS model.
Those exceptional solutions occur only at $\epsilon=n$, exactly where
the energy levels cross.
The corresponding eigenstates are therefore doubly degenerate and have
no definite parity.
They can be expressed as a product of a polynomial of finite order and
an exponential function in the Bargmann representation
$\psi(z) \propto e^{-\nu z} \prod_{i=1}^{n}(z-z_{i})$,
where the zeros $z_{i}$ are given by the roots of the Bethe ansatz
equation~(\ref{BAeq}).
Using numerical diagonalization we plot the spectrum of the
generalized Rabi model in Fig.~\ref{fig:levels} for a range of
coupling parameters.
These calculations fully confirm our expectations
about the number and positions of exceptional points, the energy level
crossings.
Below (see Sec.~\ref{subsec:examples}) we analyze several examples of
the solution in more detail.
The number of exceptional solutions for a given integer energy
$\epsilon=n$, is determined by the number of real solutions of the polynomial
equation obtained by solving the Bethe ansatz equations~(\ref{BAeq}), $p_{n}(\kappa,~\nu,~\delta)=0$
(see Appendix~\ref{appendix:C}).
In Fig.~\ref{fig:numbcross} we plot the number of exceptional
solutions, $N_{cr}$, for the first eight integer energies
$n=0,1,\ldots, 7$ as a function of $\omega_{0}$ and for fixed photon
frequency $\omega=1$ and coupling $g_{2}=0.01$.
We find that $N_{cr}$ is always between $n+1$ and $2n+1$.
In Sec.~\ref{subsec:numcrossandavoidedcross} we use a degenerate
perturbation theory to show that $N_{cr}$ depends on the detuning
$\abs{\omega-2\omega_{0}}$ and that $n+1 \leq N_{cr} \leq 2n+1$.
\begin{figure}[h!]
  \begin{center}
    \includegraphics[scale=0.83]{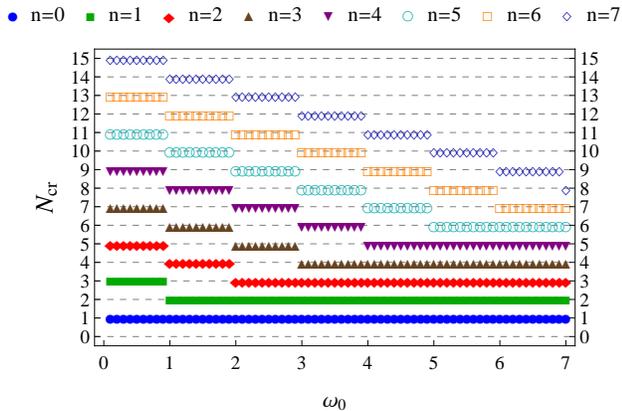}
    \caption{Number of energy level crossings, $N_{cr}$, for a given integer
      energy $n$ as a function of $\omega_{0}$ for $\omega=1$ and $g_{2}=0.01$.}
    \label{fig:numbcross}
  \end{center}
\end{figure}
%


\subsubsection{Examples}
\label{subsec:examples}

\begin{figure}[h!]
  \begin{center}
    \psfrag{x}{$a$}
    \includegraphics[scale=0.83]{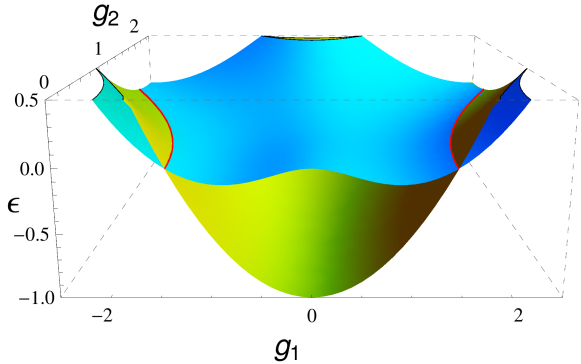}
    \caption{Plot of the lowest two energies of
      $\op{H}_{\mathrm{gR}}/\omega$ shifted by the constant $(g_{1}^{2}+g_{2}^{2})/(2\omega^{2})$,
      as a function of $g_{1}$ and $g_{2}$ for $\omega=\omega_{0}=1$.
      The energies were calculated by numerical diagonalization. 
      The bosonic Hilbert space $\{|n\rangle\}$
      was truncated by $n_{\mathrm{max}}=200$. 
      The yellow plane corresponds to the lowest energy level and the
      cyan plane to the energy of the first excited state. 
      They cross exactly in the plane of $\epsilon=0$. The lines on which
      these energy levels cross is given by
      $g_{1}^{2}-g_{2}^{2}=2\omega\omega_{0}$ (red line) as predicted by the
      quasi-exact solutions.
      }
    \label{fig:grs12}
  \end{center}
\end{figure}

Let us first consider $\epsilon=n=0$.
In this case the Bethe ansatz equations are degenerate, $Z_{1}=Z_{2}=0$ 
and the two conditions Eq.~(\ref{txfz1-z2}) and Eq.~(\ref{txsz1-z2})
are satisfied simultaneously as soon as $\kappa=\nu$, that is when
$\lambda_{-}=\delta$. 
In Fig.~\ref{fig:grs12} we plot the two lowest eigenenergies $\epsilon$ of
$\op{H}_{\mathrm{gR}}/\omega+(g_{1}^{2}+g_{2}^{2})/2\omega^{2}$ as a
function of $g_{1}$ and $g_{2}$. 
They cross precisely in the plane $\epsilon=0$ and on the curves
$\lambda_{-}=\delta$ (bold red line). In terms of the original
parameters these curves are given by
$g_{1}^{2}-g_{2}^{2}=2\,\omega\,\omega_{0}$.
The corresponding eigenstates are
$\ket{\psi_{0}}=\frac{1}{\sqrt{2}}(\ket{\nu}\ket{+} \pm \ket{-\nu}\ket{-})$, 
so-called cat states~\cite{brune1992},
where $\ket{\nu}=\exp(-\nu^{2}/2)\exp(\nu\op{a}^{\dag})\ket{0}$ is a coherent state with
$\nu=\sqrt{g_{1}g_{2}}/\omega$ and $\ket{\pm}$ are the eigenstates of $\op{\sigma}_{z}$.
%
%

For $\epsilon=n=1$ the Bethe ansatz equations~(\ref{BAeq}) can be
solved analytically
\beq
z_{1,\pm}=\frac{\kappa \nu - \nu^{2} - 1 \pm \sqrt{\nu^{2} (\kappa + \nu)^{2} + 1}}{2 \nu }. 
\eeq
The locations of the exceptional solutions in parameter space are
obtained as follows.
By inserting this expression for the Bethe root $z_{1,\pm}$ into
the two conditions Eq.~(\ref{txfz1-z2}) and Eq.~(\ref{txsz1-z2}),
we get a polynomial equation for the parameters
$\kappa,~\nu$ and $\delta$, which we denote by
$p_{1}(\kappa,\nu,\delta)=0$.
The real zeros of this equation determine the positions of the exceptional
solutions in parameter space.
The values of the original parameters are obtained by inverting the
expressions for $\kappa,~\nu$ and $\delta$:
%
\beeq
  g_{1}
  &=&
  \omega \sqrt{\frac{\nu}{\kappa}} \sqrt{\delta + \sqrt{\delta^{2}+\kappa^{2}\nu^{2}}}, 
  \\
  g_{2}
  &=&
  \omega \sqrt{\frac{\kappa}{\nu}} \frac{\nu^{2}}{\sqrt{\delta + \sqrt{\delta^{2}+\kappa^{2}\nu^{2}}}},
  \\
  \omega_{0}
  &=&
  \omega \delta.
\eeeq
We note that for a given energy level crossing only one of the two
possible Bethe roots $z_{1,\pm}$ yields a real solution to the equation
$p_{1}(\kappa,\nu,\delta)=0$, which we call $z_{1}^{\ast}$.
The corresponding eigenstates are as expected doubly degenerate
$\ket{\psi_{1}}=\frac{1}{\sqrt{2}}((\op{a}^{\dag}-z_{1}^{\ast})\ket{\nu}\ket{+} \pm (\op{a}^{\dag}+z_{1}^{\ast})\ket{-\nu}\ket{-})$.

\begin{figure}[h!]
  \begin{center}
     \includegraphics[scale=0.65]{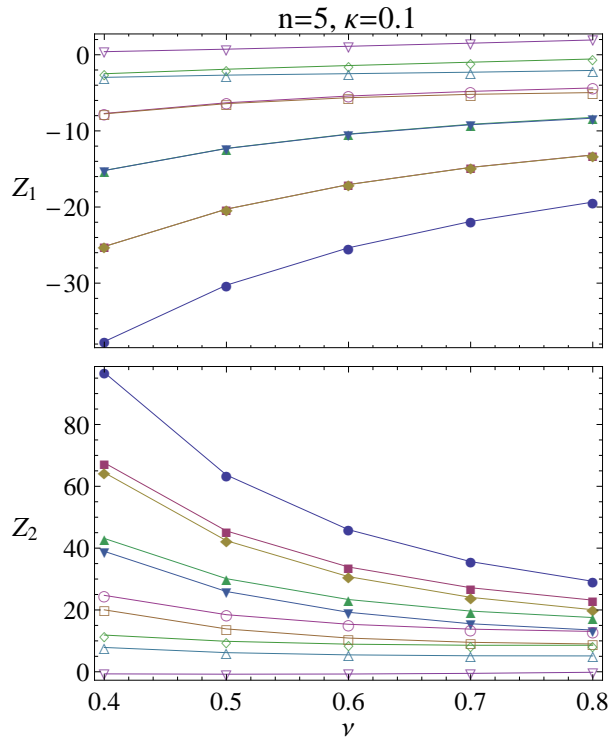}
     \caption{$Z_{1}=\sum_{i=1}^{n}z_{i}$ and
       $Z_{2}=\sum_{i=1}^{n}z_{i}^{2}$ as functions of $\nu$ for $n=5$
       and $\kappa=0.1$. Note that some lines are nearly degenerate.}
    \label{fig:z2omar}
  \end{center}
\end{figure}

For $\epsilon=n>1$ we have to solve the Bethe
equations~(\ref{BAeq}) numerically by the procedure described in
Appendix~\ref{appendix:C} to obtain the positions of the energy level
crossings in parameter space.
In Fig.~\ref{fig:z2omar} we computed the values of $Z_{1}$ and $Z_{2}$ for $n=5$ and $\kappa=0.1$ as
a function of $\nu$.
We find exactly $10$ different solutions for $Z_{1}$ and $Z_{2}$.
Inserting each corresponding pair of $Z_{1}$ and $Z_{2}$ into
Eq.~(\ref{txfz1-z2}) and Eq.~(\ref{txsz1-z2}) yields 
a polynomial equation for $\delta$.
The real zeros of this equation gives the values of $\delta$ at which
the energy levels cross on the line $\epsilon=5$.


\subsubsection{Limiting cases}

For some limits the form of the curves corresponding to exceptional
solutions can be determined analytically.
Here we use results known for the reduced BCS (Richardson) model. 
First, we rescale the roots $z_{j}=\nu x_{j}$ and rewrite~(\ref{BAeq}) as
\be
\sum_{j \neq i}^{n} \frac{2}{x_{j}-x_{i}}
  +
  \frac{\epsilon-1}{x_{i}-1}
  +
  \frac{\epsilon}{x_{i}+1}
  +
  \frac{1}{x_{i}-\frac{\kappa}{\nu}}
  +
  2\nu^{2}
  =
  0,
\ee
known as the Richardson equations in the BCS context
$g_{\mathrm{BCS}}=(2\nu^{2})^{-1}$.
The BCS pair energy levels are given by $E_{1,2}=\pm 1$ and 
the limit $\nu\rightarrow 0$ corresponds to the strong-coupling limit
in the sense of the BCS model. 
In this case the structure of the roots $\{x_{j}\}$
for the Richardson ground state solution (for which all the
Bethe roots diverge) reads~\cite{debaer}
\beq
x_{j}=\frac{1}{2\nu^{2}}y_{j}+\frac{1}{2n}\left(\frac{\kappa}{\nu}-1\right)+O(\nu^{2}),
\eeq
where $y_{j}$ are the roots of the associated Laguerre polynomials
$L^{(-1-2n)}_{n}(y)$.
Representing
$L_{n}^{(\alpha)}(y)=((-1)^{n}/n!)\prod_{j=1}^{n}(y-y_{j})$, one can
derive the sum of the roots
\begin{align}
\sum_{j=1}^{n}y_{j} &=
-n\nu\frac{d^{(n-1)}}{dy^{(n-1)}} \left. L_{n}^{(\alpha)}(y)\right|_{y=0} \nonumber \\
&= n L_{1}^{(\alpha+n-1)}(0)=n(n+\alpha),
\end{align}
as well as 
\begin{align}
\sum_{j<k}y_{j}y_{k}&=n(n-1)(n+\alpha)(n+\alpha-1)/2, \\
\sum_{j=1}^{n}y_{j}^{2}&=n(n+\alpha)(2n+\alpha-1).
\end{align}
It follows then
\begin{align}
Z_{1} &= \frac{-n(n+1)}{2\nu}+\frac{\nu}{2}(\frac{\kappa}{\nu}-1),\\
Z_{2} &= \frac{n(n+1)}{2\nu^{2}}-\frac{n+1}{2}(\frac{\kappa}{\nu}-1)+\frac{\nu^{2}}{4n}(\frac{\kappa}{\nu}-1)^{2}.
\end{align}
For the other solutions one must consider the various combinations
of diverging and non diverging roots. More details are given in~\cite{YBA}.
Interestingly, in the opposite limit of weak-coupling
$g_{\mathrm{BCS}} \rightarrow 0$ the roots can be expressed in terms of the
Laguerre polynomials (see the Refs.~\cite{YBA} and~\cite{FAG}) and the
analytical expressions for $Z_{1,2}$ can also be found.

\begin{figure}[h!]
  \begin{center}
     \includegraphics[scale=0.65]{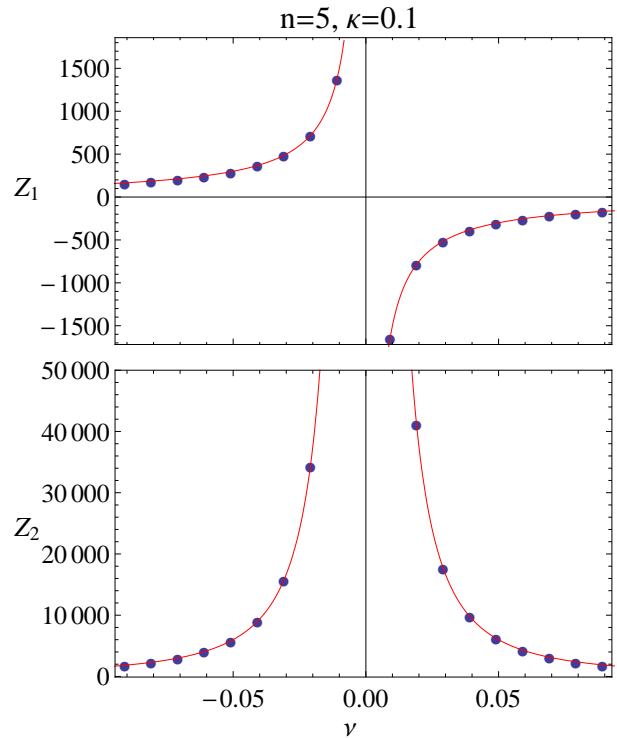}
     \caption{Comparison of the analytical limit (red line) and numerically
      calculated (blue dots) values of $Z_{1}$ and $Z_{2}$ for $n=5$,
      $\kappa=0.1$ and $\nu \to 0$.
      }
    \label{fig:z1z2limandnum}
  \end{center}
\end{figure}

In Fig.~\ref{fig:z1z2limandnum} we illustrate the analytically calculated 
limit $\nu \to 0$ of $Z_{1}$ and $Z_{2}$ (red line)
compared with some numerical values (blue dots) for $n=5$
and $\kappa=0.1$.
This is consistent with the ground state solutions of the Richardson equations.
%


\section{Regular spectrum of the generalized Rabi Hamiltonian: limits}
\label{sec:regularspectrum}

Here we consider the two limiting cases for the regular part of the
spectrum: 
(i) the limit of either small $g_{1}$ or small $g_{2}$, and 
(ii) the limit of both large $g_{1}$ and $g_{2}$.
We show that in the latter case the spectrum is a
superposition of two quasi-degenerate harmonic ladders.


\subsection{Limit of small $g_{1}$ or small $g_{2}$}

Let us focus first on the case of small $g_{2}$, $g_{2} \ll g_{1}$. 
In this limit we consider the counter-rotating part 
$\op{H}_{g_{2}}=g_{2}(\op{a}^{\dag}\,\op{\sigma}_{+}+\op{a}\,\op{\sigma}_{-})$
as a perturbation to the Jaynes-Cummings model
$\op{H}_{0} = \omega \op{a}^{\dag}\op{a} + \omega_{0}\op{\sigma}_{z} + g_{1}(\op{a}^{\dag}\,\op{\sigma}_{-} + \op{a}\,\op{\sigma}_{+})$.
For the unperturbed part $\op{H}_{0}$ we know the eigenenergies~\cite{JC}
\beq
 E_{n,k}^{(0)} &= \omega \lp n + \frac{1}{2} \rp + (-1)^{k} \Omega_{n}, 
 \\
 \Omega_{n} &= \sqrt{\lp\omega_{0}-\frac{\omega}{2}\rp^{2} + g_{1}^{2}(n+1)},
\eeq
with $k=0,1$ and the eigenstates
\beq
 \ket{n,0} &= \cos\frac{\alpha_{n}}{2} \ket{n,+} + \sin\frac{\alpha_{n}}{2} \ket{n+1,-},
 \\
 \ket{n,1} &= -\sin\frac{\alpha_{n}}{2} \ket{n,+} + \cos\frac{\alpha_{n}}{2} \ket{n+1,-},
\eeq
where
\be
  \cos\alpha_{n} = \frac{\omega_{0}-\frac{\omega}{2}}{\Omega_{n}}, 
  \qquad
  \sin\alpha_{n} = \frac{g_{1}\sqrt{n+1}}{\Omega_{n}}.
\ee
The bare basis states are defined by 
$\ket{n,\pm}=\ket{n}_{\mathrm{field}}\otimes\ket{\pm}_{\mathrm{atom}}$,
the tensor product of the Fock states $\ket{n}_{\mathrm{field}}$ and
the eigenstates of 
$\op{\sigma}_{z}$, $\op{\sigma}_{z}\ket{\pm}_{\mathrm{atom}}=\pm\ket{\pm}_{\mathrm{atom}}$.
The eigenstates $\ket{n,k}$ are simultaneously the eigenstates of the
excitation number operator, $\op{N}_{\mathrm{ex}}\ket{n,k}=(n+1)\ket{n,k}$.

The second order correction to $E_{n,k}^{(0)}$ due to
$\op{H}_{g_{2}}$ reads
%
\beq
\frac{1}{g_{2}^{2}}E_{n,0}^{(2)}
&=
-
\frac{n+2}{4\Omega_{n}}
\frac{\Omega_{n}-\omega_{0}+\frac{\omega}{2}-(n+1)\frac{g_{1}^{2}}{2\omega}}{\omega-\Omega_{n}-\frac{g_{1}^{2}}{2\omega}}
\nonumber \\
&\phantom{=}+
\frac{n}{4\Omega_{n}}
\frac{\Omega_{n}+\omega_{0}-\frac{\omega}{2}+(n+1)\frac{g_{1}^{2}}{2\omega}}{\omega+\Omega_{n}+\frac{g_{1}^{2}}{2\omega}},
\label{eq:enzerosopt}
\\
\frac{1}{g_{2}^{2}}E_{n,1}^{(2)}
&=
-
\frac{n+2}{4\Omega_{n}}
\frac{\Omega_{n}+\omega_{0}-\frac{\omega}{2}+(n+1)\frac{g_{1}^{2}}{2\omega}}{\omega+\Omega_{n}-\frac{g_{1}^{2}}{2\omega}}
\nonumber \\
&\phantom{=}+
\frac{n}{4\Omega_{n}}
\frac{\Omega_{n}-\omega_{0}+\frac{\omega}{2}-(n+1)\frac{g_{1}^{2}}{2\omega}}{\omega-\Omega_{n}+\frac{g_{1}^{2}}{2\omega}}.
\label{eq:enonesopt}
\eeq
We note that the denominators in~(\ref{eq:enzerosopt})
and~(\ref{eq:enonesopt}) can diverge.
In the following we will show that these singularities
occur at the energy levels crossings of $\op{H}_{0}$
for which the eigenenergies $E_{n,k}^{(0)} + \frac{g_{1}^{2}+g_{2}^{2}}{2\omega}$
are half-integer-valued,
and that those singularities correspond to the 
avoided level crossings in the spectrum of $\op{H}_{\mathrm{gR}}/\omega+\lambda_{+}$ at
\textit{half-integer} energies.

Let us first consider the regime $\omega\sqrt{2}>g_{1}$.
In this regime there are two singularities:
1a) $\omega=\Omega_{n}+\frac{g_{1}^{2}}{2\omega}$, which corresponds
to the degeneracy of the levels 
\be
E_{n,0}^{(0)}=E_{n+2,1}^{(0)}=\omega(n+\frac{3}{2})-\frac{g_{1}^{2}}{2\omega},
\ee
or $-2\omega+\Omega_{n}+\Omega_{n+2}=0$.
The solution of this equation is given by
\be
\frac{g_{1}^{2}}{2\omega}
=
\omega
\lb   
 n+2-\sqrt{(n+1)(n+3)+\lp\frac{1}{2}-\frac{\omega_{0}}{\omega}\rp^{2}}
\rb,
\ee
for $n=0,1,2,\ldots$.

2a) $\omega=\Omega_{n}-\frac{g_{1}^{2}}{2\omega}$, which corresponds
to the degeneracy of the levels
\be
E_{n,1}^{(0)}=E_{n-2,0}^{(0)}=\omega(n-\frac{1}{2})-\frac{g_{1}^{2}}{2\omega}, 
\ee
or $2\omega-\Omega_{n}-\Omega_{n-2}=0$.
The solution of this equation is given by
\be
\frac{g_{1}^{2}}{2\omega}
=
\omega
\lb   
 n-\sqrt{(n-1)(n+1)+\lp\frac{1}{2}-\frac{\omega_{0}}{\omega}\rp^{2}}
\rb,
\ee
for $n=2,3,4,\ldots$.

Second, we consider the regime $\omega\sqrt{2}<g_{1}$.
In this regime there are also two singularities:
1b) $\omega=-\Omega_{n}+\frac{g_{1}^{2}}{2\omega}$, which corresponds
to the degeneracy of the levels
\be
E_{n,1}^{(0)}=E_{n+2,1}^{(0)}=\omega(n+\frac{3}{2})-\frac{g_{1}^{2}}{2\omega},
\ee
or $-2\omega-\Omega_{n}+\Omega_{n+2}=0$.
The solution of this equation is given by
\be
\frac{g_{1}^{2}}{2\omega}
=
\omega
\lb
 n+2+\sqrt{(n+1)(n+3)+\lp\frac{1}{2}-\frac{\omega_{0}}{\omega}\rp^{2}}
\rb,
\ee
for $n=0,1,2,\ldots$.

2b) $\omega=\Omega_{n}-\frac{g_{1}^{2}}{2\omega}$, which corresponds
to the degeneracy of the levels
\be
E_{n,1}^{(0)}=E_{n-2,1}^{(0)}=\omega(n-\frac{1}{2})-\frac{g_{1}^{2}}{2\omega},
\ee
or $2\omega-\Omega_{n}+\Omega_{n-2}=0$.
The solution of this equation is given by
\be
\frac{g_{1}^{2}}{2\omega}
=
\omega
\lb
 n+\sqrt{(n-1)(n+1)+\lp\frac{1}{2}-\frac{\omega_{0}}{\omega}\rp^{2}}
\rb,
\ee
for $n=2,3,4,\ldots$.

0) In addition, we consider the corrections to the level
$E_{-1,1}^{(0)}$ with the eigenstate $\ket{-1,1}=\ket{0,-}$.
The second order correction due to $\op{H}_{g_{2}}$
reads
\be
\frac{1}{g_{2}} E_{-1,1}^{(2)}
=
\frac{1}{4}
\frac{1}{\omega_{0}+\frac{\omega}{2}-\frac{g_{1}^{2}}{2\omega}}.
\ee
Here the singularity can happen for
$\Omega_{1}=\omega_{0}+\frac{3}{2}\omega$, which corresponds to the
degeneracy of the levels
\be
E_{-1,1}^{(0)}=E_{1,1}^{(0)}=\frac{\omega}{2}-\frac{g_{1}^{2}}{2\omega}
\ee
The solution of this equation is given by
\be
\frac{g_{1}^{2}}{2\omega}
=
\omega_{0}+\frac{\omega}{2}.
\ee

\subsubsection{Degenerate perturbation theory}

At the degeneracy points 1a)-2b) and 0) of the unperturbed Hamiltonian
$\op{H}_{0}$, we need to use a degenerate perturbation theory to
calculate the avoided level crossings of $\op{H}_{\mathrm{gR}}$.

In the case of 1a) the gap equals to
\beq
\Delta_{n,0;n+2,1}
&= 2 \abs{\bra{n,0}\op{H}_{g_{2}}\ket{n+2,1}} 
\nonumber \\
&= 2 g_{2} \sqrt{n+2} \sin\frac{\alpha_{n}}{2}\sin\frac{\alpha_{n+2}}{2}.
\eeq
The case 2a) is obtained by the shift $n \to n-2$.
At the degeneracy of 1b) the gap equals to
\beq
\Delta_{n,1;n+2,1} 
&= 2 \abs{\bra{n,1}\op{H}_{g_{2}}\ket{n+2,1}}
\nonumber \\
&= 2 g_{2} \sqrt{n+2} \cos\frac{\alpha_{n}}{2}\sin\frac{\alpha_{n+2}}{2},
\eeq
and the case 2b) is again obtained by simply shifting $n \to n-2$.
For the crossing of $E_{-1,1}^{(0)}$ and $E_{1,1}^{(0)}$, i.e. the case 0), we have 
\beq
\Delta_{-1,1;1,1} 
&= 2 \abs{\bra{-1,1}\op{H}_{g_{2}}\ket{1,1}}
\nonumber \\
&= 2 g_{2} \sin\frac{\alpha_{1}}{2}.
\eeq

At these degeneracy points a meaningful approximation for the eigenenergies of
$\op{H}_{\mathrm{gR}}$ is given by
%
%
\beq
&E_{n,k}^{(\pm)}
=
\nonumber \\
&
\frac{E_{n,k}^{(0)}+E_{n+2,1}^{(0)} \pm \sqrt{(E_{n,k}^{(0)}-E_{n+2,1}^{(0)})^{2}+\Delta_{n,k;n+2,1}^{2}}}{2},
\label{eq:weakdegpert}
\eeq
where $(n=-1,~k=1)$ corresponds to the case 0), $(n=0,1,\ldots,~k=0)$
represents the case 1a) and $(n=0,1,\ldots,~k=1)$ gives the case 1b).
\begin{figure}[h!]
  \begin{center}
     \includegraphics[scale=0.9]{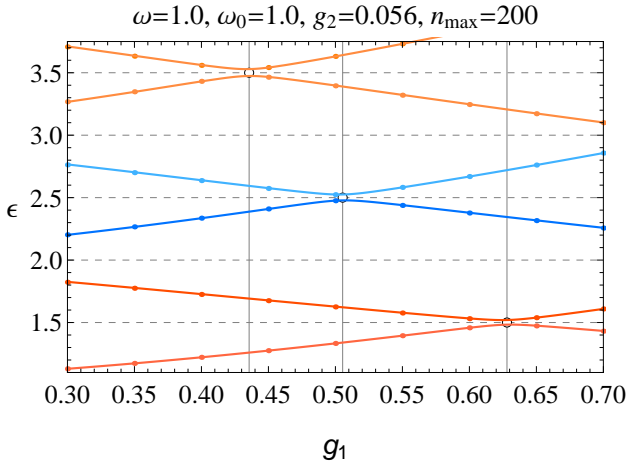}
    \caption{The weak-coupling approximation given by Eq.~(\ref{eq:weakdegpert}) for the
      spectrum of the generalized Rabi model (full lines) compared
      with the numerical calculation of the spectrum (dots) for
      $g_{2}=0.056$ and $\omega=\omega_{0}=1$.
      Note that we added the constant $\lambda_{+}=\frac{g_{1}^{2}+g_{2}^{2}}{2\omega^{2}}$ to the
      Hamiltonian $\op{H}_{\mathrm{gR}}/\omega$
      such that the level crossings occur at integer
      values and the avoided level crossings at half-integer values.
      }
    \label{fig:weakfv}
  \end{center}
\end{figure}

Other avoided level crossings can happen at $E_{n,k}^{(0)}=E_{n+2p,1}^{(0)}$ and
$E_{-1,1}^{(0)}=E_{-1+2p,1}^{(0)}$, where $k=0,1$ and $p>1$ is an
integer.
The corresponding gaps are
$\Delta_{n,k;n+2p,1}\sim\mathcal{O}(g_{2}^{p})$, since the
corresponding eigenstates can only be connected in the perturbation
theory by an application of $\op{H}_{g_{2}}$ at least $p$ times.

Let us consider $p>1$. Then
\be
\label{eq:nptwop}
\Omega_{n+2p} \pm \Omega_{n} = 2 \, p \, \omega
\ee
and
\be
\label{eq:moneptwop}
\Omega_{-1+2p} = \omega_{0} + \omega (2p-\frac{1}{2}).
\ee
Eq.~(\ref{eq:nptwop}) has the solution for the upper sign $(k=0)$ at
\beq
&\frac{g_{1}^{2}}{2\omega^{2}}
=
n+p+1-\sqrt{(n+1)(n+2p+1)+\lp\frac{1}{2}-\frac{\omega_{0}}{\omega}\rp^{2}},
\label{eq:g1pcondavcr}
\\
&
E_{n,0}^{(0)}
=
E_{n+2p,1}^{(0)}
=\omega(n+p+\frac{1}{2})-\frac{g_{1}^{2}}{2\omega},
\eeq
for $\omega\sqrt{2p}>g_{1}$.
It is also important that the rhs of~(\ref{eq:g1pcondavcr}) is greater or equal to zero,
which implies
\be
p^{2} \geq \lp \frac{1}{2} - \frac{\omega_{0}}{\omega} \rp^{2}.
\ee
The solutions for the lower sign $(k=1)$ takes place at
\beq
&
\frac{g_{1}^{2}}{2\omega^{2}}
=
n+p+1+\sqrt{(n+1)(n+2p+1)+\lp\frac{1}{2}-\frac{\omega_{0}}{\omega}\rp^{2}},
\label{eq:g1pcondavcrkone}
\\
&
E_{n,1}^{(0)}
=
E_{n+2p,1}^{(0)}
=\omega(n+p+\frac{1}{2})-\frac{g_{1}^{2}}{2\omega},
\eeq
for $\omega\sqrt{2p}<g_{1}$.
Finally, Eq.~(\ref{eq:moneptwop}) has the solution
\beq
\frac{g_{1}^{2}}{2\omega}
&=
\omega_{0}+\omega(p-\frac{1}{2}),
\\
E_{-1,1}^{(0)}
&=
E_{-1+2p,1}^{(0)}
=
\omega(p-\frac{1}{2})-\frac{g_{1}^{2}}{2\omega}.
\eeq

\subsubsection{Number of crossings and avoided crossings}
\label{subsec:numcrossandavoidedcross}

So we have seen that the avoided level crossings always happen at
\textit{half-integer} energies.
By analogous considerations one can show that the crossings always
happen at \textit{integer} energies.

To find crossing points we need to solve the equations
$E_{n,k}^{(0)}=E_{n+2p-1,1}^{(0)}$ and $E_{-1,1}^{(0)}=E_{-2+2p,1}^{(0)}$ for
$p>1$, which is equivalent to solve the equations
\beq
\Omega_{n+2p-1} \pm \Omega_{n}
&=
(2p-1) \omega
\label{eq:nbcrptone}
\\
\Omega_{-2+2p} &= \omega_{0} + \omega(2p-\frac{3}{2}).
\label{eq:nbcrpttwo}
\eeq

Eq.~(\ref{eq:nbcrptone}) has the solutions for the upper sign $(k=0)$ at
\beq
&
\frac{g_{1}^{2}}{2\omega^{2}}
=
n+p+\frac{1}{2}
-
\sqrt{(n+1)(n+2p)+\lp \frac{1}{2} - \frac{\omega_{0}}{\omega} \rp^{2}},
\label{eq:g1posp}
\\
&
E_{n,0}^{(0)}=E_{n+2p-1,1}^{(0)}=\omega(n+p)-\frac{g_{1}^{2}}{2\omega},
\label{eq:enoenp2m11}
\eeq
for $\omega\sqrt{2p-1}>g_{1}$.
It is also important that the rhs of~(\ref{eq:g1posp}) is greater or
equal to zero, which implies
\be
\lp
p-\frac{1}{2}
\rp^{2}
\geq
\lp
\frac{1}{2}
-
\frac{\omega_{0}}{\omega}
\rp^{2}
\quad
\mbox{or}
\quad
p \geq \frac{1}{2}+\left| \frac{1}{2}-\frac{\omega_{0}}{\omega}\right|.
\ee
The solution for the lower sign $(k=1)$ takes place at
\beq
&
\frac{g_{1}^{2}}{2\omega^{2}}
=
n+p+\frac{1}{2}+
\sqrt{(n+1)(n+2p)+\lp\frac{1}{2}-\frac{\omega_{0}}{\omega}\rp^{2}},
\\
&
E_{n,1}^{(0)}=E_{n+2p-1,1}^{(0)}=\omega(n+p)-\frac{g_{1}^{2}}{2\omega},
\label{eq:en1enp2pm11}
\eeq
for $\omega\sqrt{2p-1}<g_{1}$.

Eq.~(\ref{eq:nbcrpttwo}) has the solution
\beq
\frac{g_{1}^{2}}{2\omega}
&=
\omega_{0}+\omega(p-1),
\\
E_{-1,1}^{(0)}&=E_{-2+2p,1}^{(0)}=\omega(p-1)-\frac{g_{1}^{2}}{2\omega}.
\eeq

To count the number of energy level crossings at a given
integer $N$, we cast Eq.~(\ref{eq:enoenp2m11}) to
\be
E_{N-p,0}^{(0)} + \frac{g_{1}^{2}}{2\omega}
=
E_{N+p-1,1}^{(0)} + \frac{g_{1}^{2}}{2\omega}
=
\omega \, N,
\ee
with
\be
\frac{1}{2} + \left| \frac{1}{2}-\frac{\omega_{0}}{\omega} \right|
\leq
p
\leq
N,
\ee
and Eq.~(\ref{eq:en1enp2pm11}) to
\be
E_{N-p,1}^{(0)}+\frac{g_{1}^{2}}{2\omega}
=
E_{N+p-1,1}^{(0)}+\frac{g_{1}^{2}}{2\omega}
=
\omega \, N,
\ee
with
\be
1 \leq p \leq N.
\ee
In addition, there is always one intersection of the levels
\be
E_{-1,1}^{(0)}+\frac{g_{1}^{2}}{2\omega}
=
E_{2N,1}^{(0)}+\frac{g_{1}^{2}}{2\omega}
=
\omega \, N.
\ee
Thus, the number of crossings depends on the value of the detuning
$\abs{\omega-2\omega_{0}}$.
Analogously, we find the number of avoided level-crossings.

Finally, we would like to point out that the case of $g_{1} \ll g_{2}$,
can be reduced to the previous case by simply exchanging $g_{1} \leftrightarrow g_{2}$ and
flipping the sign of the level splitting $\omega_{0} \to -\omega_{0}$
in all the formulas above, since
$\op{H}_{\mathrm{gR}}(\omega,\omega_{0},g_{1},g_{2})=\op{T}^{\dag}\op{H}_{\mathrm{gR}}(\omega,-\omega_{0},g_{2},g_{1})\op{T}$
where $\op{T}=\exp(i\frac{\pi}{2}\op{\sigma}_{y})\exp(i\pi\op{a}^{\dag}\op{a})$.


\subsection{Strong-coupling limit}

When the couplings $g_{1}$ and $g_{2}$ are both large we can identify two limits. One limit corresponds to small $\omega_{0}$ the other to large $\omega_{0}$. In the former case the spectrum consists of two quasi-degenerate harmonic ladders, in the latter the spectrum is related to the solvable Jaynes-Cummings model.
%


\subsubsection{Small $\omega_{0}$ limit}
\begin{figure}[h!]
  \begin{center}
     \includegraphics[scale=0.65]{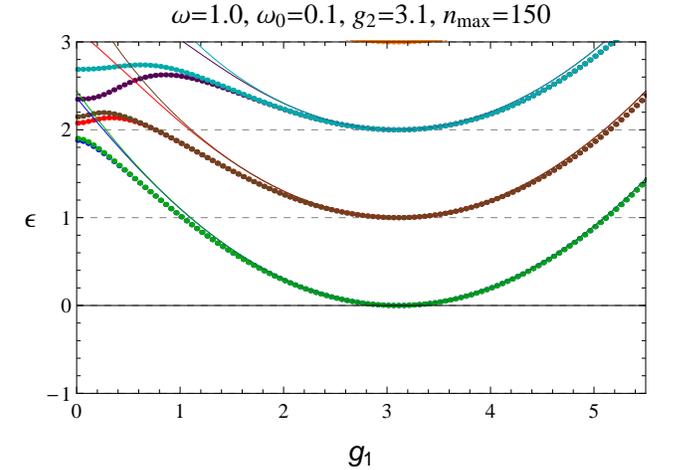}
    \caption{Comparison of the strong-coupling quasi-degenerate
             harmonic ladder structure of energy spectrum as given by 
             Eq.~(\ref{eq:strong}) (solid lines) 
             and numerical diagonalization (dots) of $\op{H}_{\mathrm{gR}}/\omega+\frac{g_{1}^{2}+g_{2}^{2}}{2\omega^{2}}$.}
    \label{fig:strong}
  \end{center}
\end{figure}

In the strong-coupling limit, where both $g_{1}$ and $g_{2}$ are large and $\omega_{0}$ is small,
one can make use of the adiabatic approximation~\cite{IGMS}.
The idea behind this approximation for the Rabi model is to rotate the
basis and to consider the term $\omega \op{a}^{\dag}\op{a}+g\op{\sigma}_{x}(\op{a}+\op{a}^{\dag})$ 
as a leading term which can be easily diagonalized, 
while the term $\omega_{0}\op{\sigma}_{z}$ is treated as a perturbation.
Generalizing this to our model we first rotate the spin basis 
$\op{\sigma}_{x} \rightarrow \op{\sigma}_{y}, \op{\sigma}_{y}
\rightarrow \op{\sigma}_{z}, \op{\sigma}_{z} \rightarrow \op{\sigma}_{x}$ 
and write
\be
 \op{H}_{\mathrm{gR}} 
 =
 \omega \op{a}^{\dag}\op{a} 
 +
 \beta( \op{a} + \op{a}^{\dag}) \op{\sigma}_{z}
 +
 i \lambda (\op{a} - \op{a}^{\dag}) \op{\sigma}_{x}
 +
 \omega_{0} \op{\sigma}_{y},
\ee 
where $\beta=(g_{1}+g_{2})/2$ and $\lambda=(g_{1}-g_{2})/2$. 
In the adiabatic approximation the terms proportional to $\omega_{0}$
and $\lambda$ should be treated as a perturbation. 
Considering the basis $|\sigma\rangle\otimes|N_{\sigma}\rangle$, 
where $\sigma=\pm$ and $|N_{\pm}\rangle = \op{D}(\mp\beta/\omega)
|N\rangle$ with the Fock states $|N\rangle$  
$(N=0,1,2,\ldots)$ and the displacement operator 
$\op{D}(\beta/\omega) = \exp\lp (\beta/\omega) ( \op{a}^{\dag} - \op{a} ) \rp$, 
we obtain the eigenvalue equation for the leading term
\begin{align}
\lb 
 \lp \op{a}^{\dag} \pm \frac{\beta}{\omega} \rp 
 \lp \op{a} \pm \frac{\beta}{\omega} \rp 
\rb \ket{\phi_{\pm}}
&\equiv
\op{D} \lp \mp \frac{\beta}{\omega} \rp \, \op{a}^{\dag} \op{a} \,
\op{D}^{\dag} \lp \mp \frac{\beta}{\omega} \rp \ket{\phi_{\pm}}
\nonumber \\
&=\lp \frac{E}{\omega} + \frac{\beta^{2}}{\omega^{2}} \rp \ket{\phi_{\pm}}.
\end{align}
In this basis the Hamiltonian approximately has a block diagonal form
with the $N$th block given by
\begin{align}
 \op{H}_{\mathrm{gR}}^{(N)} =
\begin{pmatrix}
 E_{N} & h_{-,+} \\
 h_{+,-} & E_{N}  
\end{pmatrix},
\end{align}
where 
\begin{align}
h_{-,+}
&=
-i \omega_{0} \bracket{N_{-}}{N_{+}}
+ i \lambda \bra{N_{-}}(\op{a}-\op{a}^{\dag})\ket{N_{+}}, \\
h_{+,-}
&=
i \omega_{0} \bracket{N_{+}}{N_{-}} + i \lambda \bra{N_{+}}
(\op{a}-\op{a}^{\dag}) \ket{N_{-}}, \\
E_{N}&=\omega(N-\beta^{2}/\omega^{2}),
\end{align}
provided the terms containing
the overlaps $\bracket{N_{\pm}}{M_{\mp}}$ for $N\neq M$ are
neglected.
The overlap of the two displaced coherent states is 
\be
  \bracket{M_{-}}{N_{+}}
  =
  e^{-2\beta^{2}/\omega^{2}}
  \lp \frac{2\beta}{\omega} \rp^{N-M}
  \sqrt{\frac{M!}{N!}} \, 
  L^{N-M}_{M} \lp \frac{4\beta^{2}}{\omega^{2}} \rp,
\ee
for $M<N$ and
\be
  \bracket{M_{-}}{N_{+}}
  =
  e^{-2\beta^{2}/\omega^{2}}
  \lp \frac{-2\beta}{\omega} \rp^{M-N}
  \sqrt{\frac{N!}{M!}} \,
  L^{M-N}_{N} \lp \frac{4\beta^{2}}{\omega^{2}} \rp,
\ee
for $M \geq N$,
while $\langle M_{-}|N_{+}\rangle=(-1)^{N-M}\langle
N_{-}|M_{+}\rangle$ and 
$\langle M_{+}|N_{-}\rangle=(-1)^{M-N}\langle M_{-}|N_{+}\rangle$. 
The eigenenergies of the perturbed system are
\be
 E^{\pm}_{N} = E_{N} \pm \left| \omega_{0} \bracket{N_{-}}{N_{+}} - \lambda \bra{N_{-}} (\op{a} - \op{a}^{\dag}) \ket{N_{+}} \right|.
 \ee
To compute the necessary matrix elements we use the identities 
\begin{align}
 \bracket{M_{-}}{N_{+}} &= \bra{m} \op{D}(-2\alpha) \ket{n},
 \\
 \bra{M_{-}} \op{a} \ket{N_{+}}
 &=
 \bra{m} \op{D}(-\alpha) \, \op{a} \, \op{D}(-\alpha) \ket{n}
 \nonumber \\
 &=
 \bra{m} \op{D}(-2\alpha) \op{D}(\alpha) \, \op{a} \, \op{D}(-\alpha) \ket{n}. 
\end{align}
It follows then
\begin{align}
\bra{M_{-}} \op{a} \ket{N_{+}}
&=
\bra{m} \op{D}(-2\alpha)(\op{a}-\alpha) \ket{n}
\nonumber \\
&=
\bra{m} \op{D}(-2\alpha) \lp \sqrt{N} \ket{n-1} - \alpha \ket{n} \rp
\nonumber \\
&=
\sqrt{N} \bracket{M_{-}}{(N-1)_{+}}
-
\alpha \bracket{M_{-}}{N_{+}}
\end{align}
and
\be
\bra{M_{-}} \op{a}^{\dag} \ket{N_{+}}
=
\sqrt{N+1} \bracket{M_{-}}{(N+1)_{+}} 
-
\alpha \bracket{M_{-}}{N_{+}}.
\ee
Thus we obtain the eigenenergies in the adiabatic approximation
\begin{widetext}
\be
  E_{N}^{\pm}
  =
  E_{N}
  \pm
  e^{-2\beta^{2}/\omega^{2}}
  \left|
   \omega_{0} L_{N}^{0}\lp \frac{4\beta^{2}}{\omega^{2}} \rp
   +
   \lambda \frac{2\beta}{\omega} \lb L_{N-1}^{1}\lp \frac{4\beta^{2}}{\omega^{2}} \rp + L_{N}^{1}\lp \frac{4\beta^{2}}{\omega^{2}} \rp  \rb
  \right|.
\label{eq:strong}
\ee
\end{widetext}
In the limiting case $g_{1}=g_{2}=g$ this equation agrees with the one
obtained for the Rabi model in the limit of large $g$~\cite{IGMS}.
A similar approximation scheme for the Rabi model,
using a ``symmetrizied generalized RWA''~\cite{victoralbert}, also
reproduces our result in the large $g$ limit.
We note that the second term introduces an exponentially small
splitting when $g_{1}$ and $g_{2}$ are large and therefore the spectrum is a
quasi-degenerate harmonic ladder. 
Fig.~\ref{fig:strong} illustrates the good agreement between
Eq.~(\ref{eq:strong}) and the numerical results for $\omega_{0} \ll \omega$ and
$\lambda \ll 1$.
%
%
%

\subsubsection{Large $\omega_{0}$ limit}

In this limit it is convenient to introduce the operators
\be
\op{A}=\frac{1}{g_{-}}\lp g_{1}\op{a} + g_{2}\op{a}^{\dag} \rp,
\quad
\op{A}^{\dag}=\frac{1}{g_{-}}\lp g_{1}\op{a}^{\dag} + g_{2}\op{a} \rp,
\ee
with $g_{-}\equiv\sqrt{g_{1}^{2}-g_{2}^{2}}$, such that
$\commutator{\op{A}}{\op{A}^{\dag}}=1$ holds.
We note that the operators $\op{A}$ and $\op{A}^{\dag}$ are only well
defined if $g_{1}>g_{2}$. 
The operator $\op{A}^{\dag}\op{A}$ is diagonal in the squeezed Fock
states $\ket{n,r}=\op{S}(r)\ket{n}$, where
$\op{S}(r)=\exp(\frac{1}{2}r(\op{a}^{\dag})^{2} -
\frac{1}{2}r\op{a}^{2})$ and $\{\ket{n}\}$ are the eigenstates of
$\op{a}^{\dag}\op{a}$.
So we have $\op{A}=\op{S}\,\op{a}\,\op{S}^{\dag}=\cosh(r)\op{a}+\sinh(r)\op{a}^{\dag}$
with $\tanh(r)=g_{2}/g_{1}$ and $\op{A}^{\dag}\op{A}|n,r\rangle=n|n,r\rangle$, where
$n=0,1,2,\ldots$.
Using the operators $\op{A}$ and $\op{A}^{\dag}$ we can rewrite the
Hamiltonian of the generalized Rabi model~(\ref{gR}) as
\be
\op{H}_{\mathrm{gR}}
=
\op{H}_{0}
-
\omega\frac{g_{1}g_{2}}{g_{-}^{2}}\op{H}',
\ee
where 
$\op{H}_{0}=\omega_{g} \op{A}^{\dag}\op{A}+\omega_{0}
\op{\sigma}_{z}+g_{-}( \op{A}^{\dag}\op{\sigma}_{-} +
\op{A}\op{\sigma}_{+}) + \omega\frac{g_{2}^{2}}{g_{-}^{2}}$,
with
$\omega_{g}\equiv\omega\frac{g_{1}^{2}+g_{2}^{2}}{g_{1}^{2}-g_{2}^{2}}$
and the perturbation is given by 
$\op{H}'=\op{A}^{\dag}\op{A}^{\dag}+\op{A}\op{A}$.
The Hamiltonian $\op{H}_{0}$ has the same form as the Jaynes-Cummings
Hamiltonian apart from the additional constant
$\omega\frac{g_{2}^{2}}{g_{-}^{2}}$, therefore its spectrum and
eigenstates are known~\cite{JC}.
From this it follows that the spectrum of the generalized Rabi model
in the first order in $\op{H}'$ reads
\be
E_{n,\pm}^{(0)}
=
\omega_{g}(n-\frac{1}{2})
\pm
\frac{1}{2}
\sqrt{(2\omega_{0}+\omega_{g})^{2}+4\,n\,g_{-}^{2}}
+
\omega\frac{g_{2}^{2}}{g_{-}^{2}},
\label{eq:largeomega0}
\ee
which is valid for small $\omega\frac{g_{1}g_{2}}{g_{-}^{2}}$ and
large $\omega_{0}$.
\begin{figure}[h!]
  \begin{center}
    \includegraphics[scale=0.65]{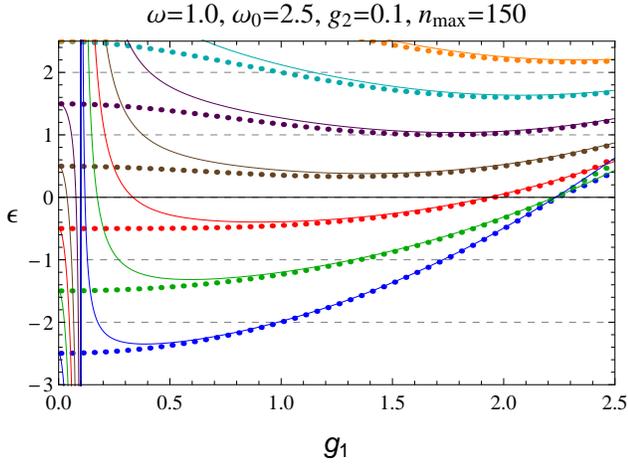}
    \caption{Comparison of the strong-coupling and large $\omega_{0}$ energy spectrum as given by 
             Eq.~(\ref{eq:largeomega0}) (solid lines) 
             and numerical diagonalization (dots) of
             $\op{H}_{\mathrm{gR}}/\omega+\frac{g_{1}^{2}+g_{2}^{2}}{2\omega^{2}}$.
             The perturbation theory~(\ref{eq:largeomega0}) clearly
             fails in the region where $g_{1} \approx g_{2}$ since
             then the expression $\omega\frac{g_{1}g_{2}}{g_{-}^{2}}$
             diverges.
             This can be seen in the plot when $g_{1}$ is equal to
             $g_{2}=0.1$, then the solid lines diverge and around this
             region the approximation~(\ref{eq:largeomega0}) is not appropriate.}
    \label{fig:largeomega0}
  \end{center}
\end{figure}

\section{Discussion and Conclusions}
\label{sec:conclusionanddiscussions}

The connection between the polynomial solutions, Bethe ansatz
equations and quasi-exact solvability is well known and has been
discussed in the literature from different perspectives, see,
e.g.,~\cite{Turbiner,zhang} and Refs. therein.
Noticing that
\be
 J^{+}=z^{2}\frac{d}{dz}-nz,\qquad J^{-}=\frac{d}{dz},\qquad J^{0}=z\frac{d}{dz}-\frac{n}{2},
\ee
is a differential realization
of the $(n+1)$-dimensional representation of the $sl(2)$ algebra in the Bargmann space, one can
construct a bilinear combination of $J^{\pm,0}$ whose eigenstates are polynomials of the order $n$ and smaller. 
This leads to a second-order differential operator which
is called quasi-exactly solvable~\cite{Turbiner}. We illustrate this
construction on the simple case of $\kappa=-\nu$ of the generalized Rabi model. 
The differential operator acting on the function $\chi(z)$ is
$(z^{2}-\nu^{2})d^{2}_{z}-(2\nu(z^{2}-\nu^{2})+2nz-2\nu)d_{z}+2\nu n z
-\mathcal{A}$, where $-\mathcal{A}=(n-2\lambda_{+})(n+1)-2\nu^{2}$ and $d_{z}\equiv
d/dz$. 
Using the operators $J^{\pm,0}$ it can be represented as
$J^{+}J^{-}-\nu^{2}J^{-}J^{-}+2\nu J^{+}+2\nu(\nu^{2}+1)J^{-}-n
J^{0}-\mathcal{A}-n^{2}/2$ which has a quasi-exactly-solvable form. This
predicts the existence of the exceptional part of the spectrum in the
generalized Rabi model which has been studied in this paper. In
particular, we found that:
(i) The exceptional part of the spectrum corresponds to the level
crossings; no level crossings occur outside of the exceptional
points.
(ii) All level crossings occur at integer values of energy
$\epsilon_{c}=n$; the number of crossings in parameter space is always
between $n+1$ and $2n+1$.
%
%
The wave functions at these points have a polynomial structure in
Bargmann space.
(iii) The avoided level crossings occur at half-integer values of the
energy, $\epsilon_{ac}=n/2$, at least for $g_{1} \gg g_{2}$ (or
$g_{2}\gg g_{1}$).
(iv) In the strong-coupling limit $g_{1}/\omega \gg 1$ and
$g_{2}/\omega \gg 1$, the spectrum consist of the two
quasi-degenerate harmonic ladders.

The obtained results for the generalized Rabi model can be used in
several physical applications, namely for the two-dimensional electron
gas in a magnetic field with Rashba and Dresselhaus spin-orbit couplings
and for the cavity and circuit QED systems.

\section{Acknowledgments}

This work was supported by the Swiss National Science
Foundation. M.T. is grateful to IIP for hospitality.

\begin{appendix}

\section{Derivation of Eq.~(\ref{eq:diffeqchisimp})}
\label{appendix:A}

In this Appendix we show how the system of two first-order
differential equations~(\ref{eq:psiztwop}) and~(\ref{eq:psizonep}) can
be reduced to a single second order differential equation for
$\psi_{1}(z)$.
To this end we differentiate Eq.~(\ref{eq:psizonep}) one more time and
eliminate $\psi_{2}(z)$ from Eq.~(\ref{eq:psiztwop}) and $\psi_{2}'(z)$
from Eq.~(\ref{eq:psizonep}).
Thus, we get a second order differential equation for $\psi_{1}(z)$. 
The substitution 
\be
 \psi_{1}(z) = \exp(-\nu z)\chi(z),
\ee
yields the following differential equation for $\chi(z)$
\be
 A_{3}(z) \chi'' + B_{3}(z) \chi' + C_{2}(z) \chi = 0,
 \label{eq:diffeqchiapB}
\ee
where the polynomials are
\be
 A_{3}(z) = \sum_{j=0}^{3} a_{n} z^{n},
 \quad
 B_{3}(z) = \sum_{j=0}^{3} b_{n} z^{n},
 \quad
 C_{2}(z) = \sum_{j=0}^{2} c_{n} z^{n},
\ee
with the corresponding coefficients given by
\begin{align}
 a_{0} &= - \nu^{2} \delta,
 \qquad
 a_{1} =  \nu \lambda_{-},
 \nonumber \\
 a_{2} &= \delta,
 \qquad
 a_{3} = - \frac{1}{\nu} \lambda_{-},
 \\
 b_{0} &= \nu (\delta - \lambda_{-} + 2 \nu^{2}\delta),
 \quad
 b_{1} =  \lp \delta - \lambda_{-} - 2 \delta \epsilon - 2
 \nu^{2} \lambda_{-} \rp,
 \nonumber \\
 b_{2} &=  \frac{2}{\nu} \epsilon \lambda_{-} - 2 \nu \delta,
 \qquad
 b_{3} =  2 \lambda_{-},
 \\
 c_{0} &= 
 - \lb \delta \lp \delta^{2} - e^{2} + \lambda_{+} \rp + e \lambda_{-}\rb
 -  \nu^{2} (\delta-\lambda_{-})
 -  \nu^{4} \delta,
 \nonumber \\
 c_{1} &=
 \frac{1}{\nu} 
 \lb \delta^{2}\lambda_{-} - \delta\lambda_{+} - e(e+1)\lambda_{-} \rb
 \nonumber \\
 &\phantom{=}-  \nu \lp \delta - \lambda_{-} - 2 \delta \epsilon \rp
 + \nu^{3} \lambda_{-},
 \nonumber \\
 c_{2} &= -  2 \lambda_{-} \epsilon.
\end{align}
It is convenient to rewrite Eq.~(\ref{eq:diffeqchiapB}) as
\be
 \lb 
  \frac{d^{2}}{dz^{2}} 
  + 
  \lp
   \sum_{s=1}^{3}\frac{\nu_{s}}{z-\rho_{s}} + \nu_{0} 
  \rp \frac{d}{dz}
  + 
  \frac{D_{2}(z)}{\prod_{s=1}^{3}(z-\rho_{s})} 
 \rb \chi(z) = 0,
 \label{eq:diffeqchisimpapB}
\ee
where $D_{2}(z)=\sum_{s=0}^{2}d_{s}z^{s}$ is a polynomial of degree 2
with the coefficients given by $d_{j}=c_{j}/a_{3}$, $\rho_{j}$ are the
zeros of $A_{3}(z)$ and we set $\frac{\delta \nu}{\lambda_{-}} \equiv
\kappa$:
\begin{align}
  d_{0}
  &=
  \kappa
  \lp
     \delta^{2}
   - \epsilon^{2}
   + 2 \epsilon \lambda_{+}
   - \lambda_{+}^{2}
   + \lambda_{+}
   + \nu^{2}
   + \nu^{4}
  \rp
  \nonumber \\
  &+ \nu \lp \epsilon-\lambda_{+}-\nu^{2}\rp,
  \\
  d_{1}
  &=
  e(e+1)-\delta^{2}
  + 
  \delta \frac{\lambda_{+}}{\lambda_{-}}
  +
  \nu \kappa
  -
  \nu^{2}
  -
  2 \nu \epsilon \kappa
  -
  \nu^{4},\\
  d_{2} &= 2 \nu \epsilon,
  \\
  \rho_{1} &= \nu,
  \qquad
  \rho_{2} = -\nu,
  \qquad
  \rho_{3} = \kappa,
  \\
  \nu_{1} &= -\epsilon + 1,
  \quad
  \nu_{2} = -\epsilon,
  \quad
  \nu_{3} = -1,
  \quad
  \nu_{0} = -2 \nu.
\end{align}

\section{Derivation of Eq.~(\ref{txfz1-z2}) and  Eq.~(\ref{txsz1-z2})}
\label{appendix:B}

Note that we have the identities
\begin{align}
& \sum_{s=1}^{3} \rho_{s} = \kappa, \qquad
  \sum_{s=1}^{3} \nu_{s} = - 2 \epsilon, \qquad
  \sum_{s<p}^{3} \rho_{s}\rho_{p} = - \nu^{2}, \\
& \sum_{s \neq p \neq q}^{3} \nu_{s}(\rho_{p}+\rho_{q}) = -2 \epsilon \kappa - \nu + \kappa.
\end{align}
As already mentioned in the main text the first condition, Eq.~(\ref{1cond}), $d_{2}=2\nu\epsilon = 2 \nu n$, 
provides us the allowed energy spectrum,
\be
  \epsilon = n, \quad \mbox{or} \quad E = \omega (n-\lambda_{+}). 
  \label{eq:energy-exceptB}
\ee
Substituting this into the second and the third conditions, 
Eqs.~(\ref{2cond}),~(\ref{3cond}), gives two quadratic equations for
$\lambda_{+}$ and $\delta$,
\begin{align}
  2 \nu Z_{1}
  &=
  \lambda_{+}^{2}
  -
  \lp 2n+1-\frac{\kappa}{\nu} \rp \lambda_{+}
  -
  \lp \delta^{2} + \nu (\nu-\kappa)+\nu^{4} \rp, 
  \label{fz1-z2} \\
  2 \nu^{2} Z_{2}
  &=
  - \lambda_{+}^{2}
  + (2n + 1 - \frac{\kappa}{\nu} + \kappa^{2} - \nu^{2}) \lambda_{+}
  \nonumber \\
  &+ (\delta^{2} + \nu (\nu  - \kappa) + \kappa^{2}\nu^{2} + 2n \nu^{2}(\nu^{2}+1)).
 \label{sz1-z2}
\end{align}
Using the identity $\lambda_{+}^{2}-\lambda_{-}^{2}=\nu^{4}$ and that
$\lambda_{-}=\delta\nu/\kappa$ we can express $\lambda_{+}$ in terms
of $\delta$ as
$\lambda_{+}=\sqrt{\delta^{2}\frac{\nu^{2}}{\kappa^{2}}+\nu^{4}}$. 

\section{Procedure for solving the Bethe ansatz equations~(\ref{BAeq})}
\label{appendix:C}

In this appendix we present in detail the procedure for determining
the values of the parameters $\omega,~\omega_{0},~g_{1},~g_{2}$ at
which the eigenfunctions $\chi(z)$ are polynomials of finite order.
At those points the energy levels cross and
the eigenstates are doubly degenerated.
As described in the main text to determine those parameters
we have to solve the Bethe ansatz equations~(\ref{BAeq}) regarding the
conditions given by Eqs.~(\ref{txfz1-z2}),~(\ref{txsz1-z2}) and
$\epsilon=n$.

Let us consider the Bethe ansatz equations 
\beq
\sum_{j\ne i}^n \frac{2}{z_j-z_i} + \frac{\epsilon-1}{z_i-\nu} + \frac{\epsilon}{z_i+\nu} + \frac{1}{z_i-\kappa} + 2\nu = 0,
\label{apcbaeq}
\eeq
with $i=1, 2, \ldots, n$ and where we have to fix $\epsilon=n$.

These equations essentially correspond to the Bethe ansatz equations
which allow (through their solutions) to define the eigenstates of a
Reduced BCS (or Richardson) Hamiltonian~\cite{rich,rich2}.
In fact, by introducing the notation
$\epsilon_1=\nu,~\epsilon_2=-\nu,~\epsilon_3=\kappa$ and
$d_1=n-1,~d_2=n,~d_3=1$ we can write the corresponding Richardson equations in the form
\beq
r_i:=\sum_{j\ne i}^n \frac{2}{z_j-z_i}+\sum_{j=1}^3\frac{d_j}{z_i-\epsilon_j}+2\nu=0.
\label{rieq}
\eeq
Furthermore, it has been shown~\cite{FAG} that introducing the change of variables
\beq
\Lambda_j=\frac{1}{2\nu}\sum_{k=1}^n\frac{1}{\epsilon_j-z_k}
\label{cov}
\eeq
the quadratic equation
\beq
(1-d_j)\Lambda_j^{(1)}+\Lambda_j^2-\Lambda_j-\frac{1}{2\nu}\sum_{i\ne j}^3 d_i\frac{\Lambda_i-\Lambda_j}{\epsilon_i-\epsilon_j}=0
\label{quad}
\eeq
together with its derivatives
\beq
\mathcal{E}_j^{(l)}
:=&
(1-\frac{d_j}{l+1})\Lambda_j^{(l+1)}
+
\sum_{k=0}^l\binom{l}{k}\Lambda_j^{(k)}\Lambda_j^{(l-k)}\nonumber\\
&-\Lambda_j^{(l)}-l!\sum_{i\ne j}^3 d_i\left(\frac{1}{(2\nu)^{l+1}}\frac{\Lambda_i-\Lambda_j}{(\epsilon_i-\epsilon_j)^{l+1}}\right.\nonumber\\
&\left.-\sum_{m=1}^l\frac{1}{(2\nu)^m}\frac{\Lambda_j^{(l-m+1)}}{(l-m+1)!}\frac{1}{(\epsilon_i-\epsilon_j)^m}\right)=0
\label{quadder}
\eeq
for $j=1,2,3$ and $l=1,\dots,d_j-1$ form a closed system of equations
which is satisfied whenever the rapidities $z_k$ satisfy the
Richardson equations~(\ref{rieq}).

Consider now Eq.~(\ref{rieq}), by using the following relations
\beq
&\sum_{i=1}^n\sum_{j\ne i}^n \frac{1}{z_i-z_j}=0,\nonumber\\
&\sum_{i=1}^n\sum_{j\ne i}^n \frac{z_i}{z_i-z_j}=\frac{n(n-1)}{2},\nonumber\\
&\sum_{i=1}^n\sum_{j\ne i}^n \frac{z_i^2}{z_i-z_j}=(n-1)\sum_{i=1}^n z_i
\eeq
and by taking the sums $\sum_{i=1}^{n} r_i,~\sum_{i=1}^{n} r_i z_i,~\sum_{i=1}^{n} z_i^2 r_i$ we obtain
\beq
&n-\sum_{l=1}^3 d_l \Lambda_l=0,\nonumber\\
&-n(n-1) + n \sum_{l=1}^{3} d_{l} - 2 \nu \sum_{l=1}^{3} d_{l} \epsilon_{l} \Lambda_{l} + 2\nu Z_{1} = 0,\nonumber\\
&-2(n-1)Z_{1} + \sum_{l=1}^{3}d_{l} Z_{1} + n
\sum_{l=1}^{3}d_{l}\epsilon_{l} - 2 \nu
\sum_{l=1}^{3}d_{l}\epsilon_{l}^{2}\Lambda_{l} \nonumber \\
&+ 2\nu Z_{2} = 0.
\label{eqlin}
\eeq
To derive these equations we used the change of variables introduced in
Eq.~(\ref{cov}). Eq.~(\ref{eqlin}) is linear in the variables
$\Lambda_{1,2,3}$ and can be readily solved, we find
\beq
&\Lambda_{1}=
\nonumber\\
&
\frac{-n^2 (\kappa+\nu)+2 \nu  n (\kappa\nu -1)-2 Z_1 \left(\kappa  \nu +\nu ^2-1\right)+2 \nu Z_2}{4 \nu ^2 (\kappa +\nu )},\nonumber\\
&\Lambda_{2}=\frac{n^2 (\nu -\kappa )-2 \kappa  \nu ^2 n+Z_1(-2 \kappa  \nu+2 \nu ^2+2)+2 \nu  Z_2}{4 \nu ^2 (n-1) (\nu -\kappa)},\nonumber\\
&\Lambda_{3}=\frac{\kappa  n-2 \nu ^3 n-\nu  n+2 Z_1+2 \nu  Z_2}{2 \kappa ^2 \nu -2 \nu ^3}.
\eeq
Finally by solving Eq.~(\ref{quad}), with $j=1$, for $\Lambda_1^{(1)}$
and each successive derivative (Eq.~(\ref{quadder})) for
$\Lambda_1^{(l)}$ (with $l=1,\dots,d_1-1$) and then replacing $Z_{1}$
and $Z_{2}$ by the expressions given in Eq.~(\ref{txfz1-z2}) and
Eq.~(\ref{txsz1-z2}), we get at last a polynomial equation as a
function of the parameters $\kappa,~\nu$ and $\delta$ (since the first
term containing the next derivative will cancel due to the prefactor
$1-d_1/(d_1-1+1)=0$).
Let us denote this equation by $p_{n}(\kappa,~\nu,~\delta)$.
The zeros of this polynomial equation, $p_{n}(\kappa,~\nu,~\delta)=0$,
will at the end determine the position of the energy level crossings.
%
This procedure can be used to determine the positions of the crossings for
all $n>1$, for the specific case $n=1$, the Bethe ansatz
equations~(\ref{apcbaeq}) can be easily solved
\beq
z_{1}^{(\pm)}=\frac{\kappa \nu - \nu^{2} - 1 \pm \sqrt{\nu^{2} (\kappa + \nu)^{2} + 1}}{2 \nu }. 
\eeq
%


It is clear that because of the particular form of $\nu$ and $\kappa$
as functions of $g_1$ and $g_2$
\beq
\nu=\frac{\sqrt{g_1 g_2}}{\omega},~~~~\kappa=\frac{2 \delta \omega \sqrt{g_1 g_2}}{g_1^2-g_2^2},~~~~\delta=\frac{\omega_0}{\omega},
\eeq
a singularity will appear whenever we will be in the region $g_1\sim
g_2$.
Therefore, the particular case $g_1=g_2=g$ has to be treated
separately.

\subsection{Rabi limit: $g_1=g_2=g$}

In this case the corresponding Schr\"odinger equation reads
\beq
A_{2}(z) \chi''(z) + B_{2}(z) \chi'(z) + C_{1}(z)\chi(z) = 0,
\label{eq:seqrl}
\eeq
where
\beq
A_{2}(z) &= \delta (z-\nu) (z+\nu), \\
B_{2}(z) &= \delta \nu (1+2\nu^{2}) + \delta (1-2\epsilon)z  - 2\nu \delta z^{2}, \\
C_{1}(z) &= -\delta (\nu^{4}+\nu^{2}+\delta^{2}-e^{2}+\lambda_{+}) - (\frac{\delta}{\nu}\lambda_{+})z.
\eeq
By noticing that for $g_1=g_2=g$ we have $\lambda_{+}=\nu^{2}$ and
dividing the Schr\"odinger equation~(\ref{eq:seqrl}) by $A_{2}(z)$ we get
%
\beq
\bigg[ 
&\frac{d^{2}}{dz^{2}} 
 +
 \lp \frac{1-\epsilon}{z-\nu} + \frac{-\epsilon}{z+\nu} - 2\nu \rp \frac{d}{dz}
 \nonumber \\
&-
 \frac{2\nu^{2}(1+\epsilon)+\delta^{2}-\epsilon^{2} + 2\nu(1-\epsilon)z}{(z-\nu)(z+\nu)}
\bigg]
\chi(z) = 0.
\eeq
According to the Eqs.~(A.12)-(A.15) in~\cite{zhang} this differential equation has a
polynomial solution $\chi(z)=\prod_{i=1}^{n}(z-z_{i})$ of degree $n$ if 
\beq
 \label{eq:rabicondone}
 - 2 \nu(1-\epsilon)
 &=
 -n(-2\nu), \\
 \label{eq:rabicondtwo}
 -(2\nu^{2}(1+\epsilon) + \delta^{2} - \epsilon^{2})
 &=
 2\nu Z_{1} - n(n-1) - n(1-2\epsilon),
\eeq
where $Z_{1}=\sum_{i=1}^{n}z_{i}$ and $z_{i}$ are given by the roots of the
following, now much simpler, Bethe ansatz equations
\beq
  \label{eq:rabibaeq}
  \sum_{j \neq i}^{n} \frac{2}{z_{j}-z_{i}}
  +
  \frac{\epsilon-1}{z_{i}-\nu}
  +
  \frac{\epsilon}{z+\nu}
  +
  2\nu
  =
  0.
\eeq

A similar procedure to determine the values of the parameters for
which the eigenfunctions $\chi(z)$ are polynomial, and where 
the energy levels cross, can once again be applied. With the only
difference that, because of the missing term in Eq.~(\ref{eq:rabibaeq}) 
%
\beq
 \frac{1}{z_i-\kappa} \rightarrow 0,
\eeq
we will get a linear system of only two equations.
This is consistent since the condition for $Z_2$ does not apply
anymore in this case.
In view of Eq.~(\ref{eq:rabicondone}) and Eq.~(\ref{eq:rabicondtwo}), we now have to satisfy the condition
\beq
2\nu Z_1=-2\nu^{2}(n+2)-\delta^2+1
\eeq
and we have to fix $\epsilon=n+1$ in Eq.~(\ref{eq:rabibaeq}).
Thus we obtain 
\beq
\Lambda_{1} &= \frac{2 \nu Z_{1} + n \left( n + 2 + 2\nu^{2} \right)}{4 \nu^2 n}, \nonumber \\
\Lambda_{2} &= -\frac{2 \nu Z_{1} + n \left( n + 2 - 2\nu^{2} \right)}{4 \nu^2 (n+1)}.
\eeq

\begin{figure}[h!]
  \begin{center}
    \includegraphics[scale=0.65]{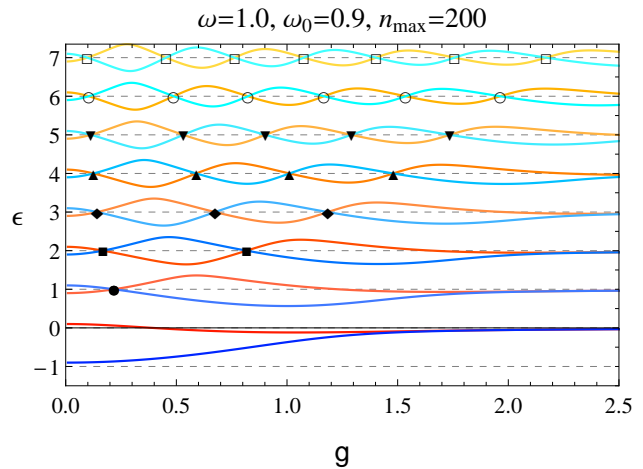}
    \caption{Plot of the energy spectrum of the Rabi model $(g_{1}=g_{2}=g)$
      obtained by numerical diagonalization. The energy level
      crossings are indicated by the black markers, which were
      calculated using the method outlined in this Appendix.
      }
    \label{fig:rabicross}
  \end{center}
\end{figure}

In Fig.~\ref{fig:rabicross} we show the crossings of the energy levels
in the Rabi limit, $g_{1}=g_{2}=g$. The energy spectrum was calculated
numerically, by truncating the bosonic Hilbert space at
$n_{\mathrm{max}}=200$. The level crossings were obtained using the
method explained above and are indicated by the black markers.

\end{appendix}


\newpage

\onecolumngrid

\begin{figure}[h!]
  \begin{center}
     \includegraphics[scale=1.0]{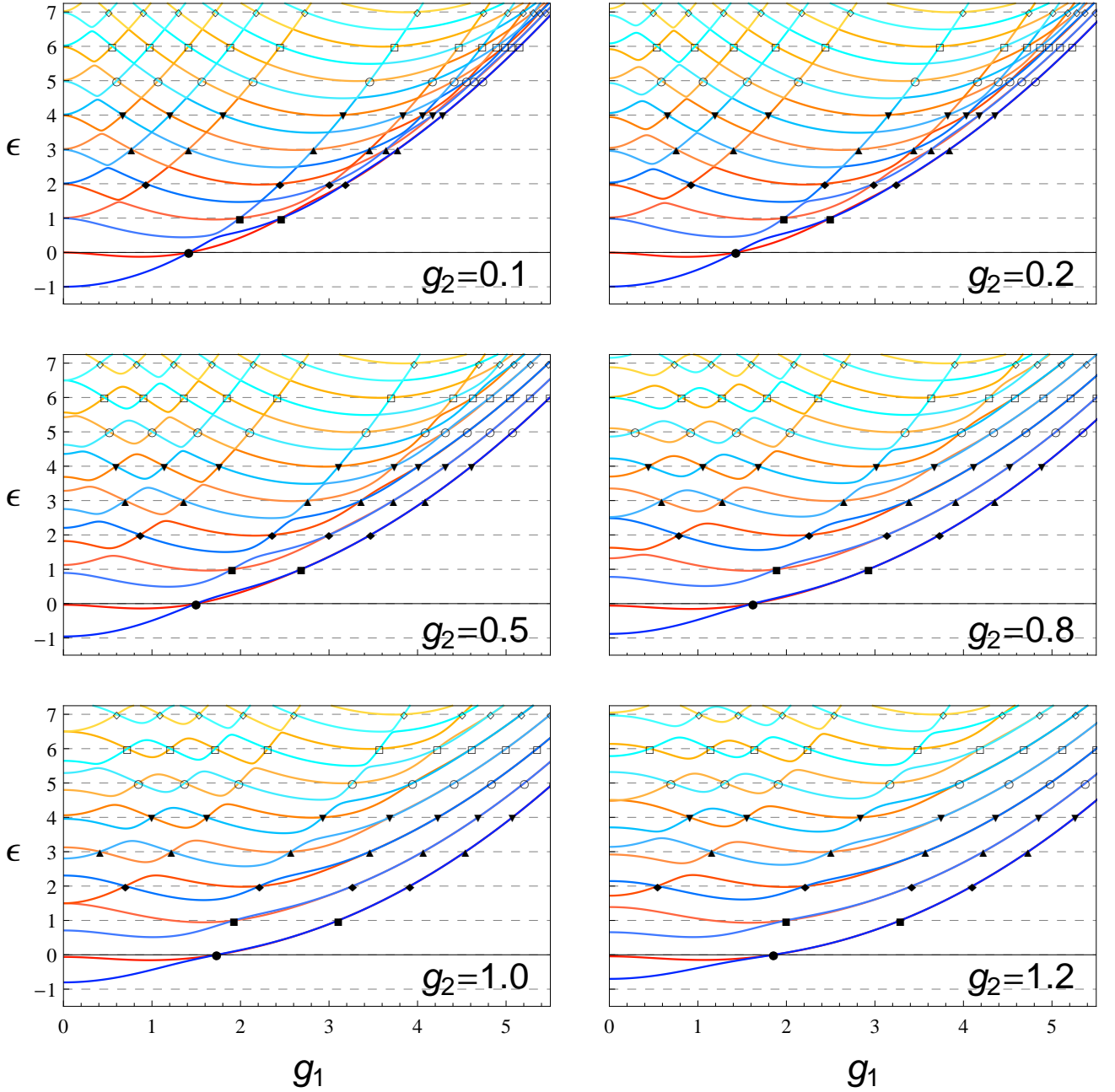}
    \caption{Energy spectrum of the generalized Rabi model shifted
      by $(g_{1}^{2}+g_{2}^{2})/(2\omega^{2})$ as a function of the
      coupling $g_{1}$ for a range of couplings $g_{2}$. 
      Black markers at integer energies $\epsilon=n$ indicate the energy
      levels intersection points, where the model has an exceptional
      spectrum.
      No other level crossings occur at different points. 
      The level repulsion happens at the half-integer values of energy.}
    \label{fig:levels}
  \end{center}
\end{figure}


\begin{thebibliography}{99}

\bibitem{Rabi}
I. I. Rabi, Phys. Rev. {\bf 49}, 324 (1936); {\bf 51}, 652 (1937).

\bibitem{JC}
E. T. Jaynes and F. W. Cummings, Proc. Inst. Elect. Eng. {\bf 51}, 89 (1963); F. W. Cummings, Phys. Rev. {\bf 140}, A1051 (1965). 


\bibitem{Braak}
D. Braak, Phys. Rev. Lett. {\bf 107}, 100401 (2011).


\bibitem{Morozov}
A. Moroz, arXiv:1205.3139 (2012); Europhys. Lett. {\bf 100}, 60010
(2012); Ann. Phys. (N.Y.) 338, 319-340 (2013).



\bibitem{Beaudoin}
F. Beaudoin, J. M. Gambetta, and A. Blais, Phys. Rev. A {\bf 84}, 043832 (2011).



\bibitem{cav-QED}
S. Haroche and J. M. Raimond, {\it Exploring the Quantum: Atoms, Caviries and photons}, (Oxford, Oxford University Press, 2006).

\bibitem{circ-QED-1}
D. I. Schuster {\it et al.}, Nature {\bf 445} 515 (2007); M. Hofheinz {\it et al.}, Nature {\bf 459} 546 (2009).

\bibitem{circ-QED-2}
P. Forn-Diaz {\it et al.}, Phys. Rev. Lett. {\bf 105} 237001 (2010); T. Niemczyk {\it et al.}, Nature Phys. {\bf 6}, 772 (2010).





\bibitem{FKU}
I. D. Feranchuk, L. I. Komarov, and A. P. Ulyanenkov, J. Phys. A: Math. Gen. {\bf 29}, 4035 (1996).

\bibitem{BS-regime}
J. Hausinger and M. Grifoni, New J. Phys. {\bf 10}, 115015 (2008).

\bibitem{IGMS}
E. K. Irish, J. Gea-Banacloche, I. Martin, and K. C. Schwab, Phys. Rev. B {\bf 72}, 195410 (2005).

\bibitem{casanova}
J. Casanova, G. Romero, I. Lizuain, J. J. Garcia-Ripoll, and E. Solano, Phys. Rev. Lett. {\bf 105}, 263603 (2010).


\bibitem{Judd}
B. R. Judd, J. Phys. C {\bf 12}, 1685 (1979).

\bibitem{Reik}
H. G. Reik, H. Nusser, and L. A. Amarante Ribeiro, J. Phys. A {\bf 15}, 3491 (1982). 

\bibitem{Kus}
M. Ku\'{s}, J. Math. Phys. {\bf 26}, 2792 (1985).

\bibitem{KL}
M. Ku\'{s} and M. Lewenstein, J. Phys. A: Math. Gen. {\bf 19}, 305 (1986).

\bibitem{RD}
H. G. Reik and M. Doucha, Phys. Rev. Lett. {\bf 57}, 787 (1986).

\bibitem{KKT}
R. Koc, M. Koca, and H. T\"{u}t\"{u}nc\"{u}ler, J. Phys. A: Math. Gen. {\bf 35}, 9425 (2002).

\bibitem{EmaryBishop}
C. Emary and R. F. Bishop, J. Math. Phys. 43, 3916 (2002).



\bibitem{Turbiner}
A. V. Turbiner, arXiv:hep-th/9409068.

\bibitem{chaos}
M. Jele\'{n}ska-Kuklinska and M. Ku\'{s}, Phys. Rev. A {\bf 41}, 2889 (1990).

\bibitem{EES}
S. I. Erlingsson, J. C. Egues, and D. Loss, Phys. Rev. B {\bf 82}, 155456 (2010).



\bibitem{SBOT}
M. Schir\'{o}, M. Bordyuh, B. \"{O}ztop, and H. E. T\"{u}reci, Phys. Rev. Lett.  {\bf 109}, 053601 (2012). 


\bibitem{GriPar2013}
A. L. Grimsmo and S. Parkins, Phys. Rev. A {\bf 87}, 033814 (2013).



\bibitem{zhang} 
Y.-Z. Zhang, J. Phys. A.: Math Theor. {\bf 45}, 065206 (2012).


\bibitem{sklyanin}
E. K. Sklyanin, Zap. nauch. semin. LOMI {\bf 134}, 112 (1983). 



\bibitem{FAG}
A. Faribault, O. El Araby, C. Str\"{a}ter, and V. Gritsev, Phys. Rev. B {\bf 83}, 235124 (2011); O. El Araby, V. Gritsev, and A. Faribault, Phys. Rev. B {\bf 85}, 115130 (2012).


\bibitem{Links}
I. Marquette and J. Links, J. Stat. Mech. (2012) P08019; F. Pan {\it et al.}, J. Phys. A: Math. Theor. {\bf 44}, 395305 (2011).

\bibitem{genGaud}
J. Dukelsky, S. Pittel, and G. Sierra, Rev. Mod. Phys. {\bf 76}, 643 (2004).

\bibitem{Ismail}
M. E. H. Ismail, Pacific J. Math.
{\bf 193}, 335 (2000); M. E. H. Ismail in {\it Random Matrix models and Their Applications}, (Cambridge University Press, 2001).





\bibitem{ML}
D. Mattis and E. Lieb, J. Math. Phys. {\bf 2}, 602 (1961).


\bibitem{brune1992}
M. Brune, S. Haroche, J. M. Raimond, L. Davidovich, and N. Zagury, Phys. Rev. A {\bf 45}, 5193 (1992).

\bibitem{debaer}
S. De Baerdemacker, Phys. Rev. C {\bf 86}, 044332 (2012).

\bibitem{YBA}
E. A. Yuzbashyan, A. A. Baytin, and B. L. Altshuler, Phys. Rev. B {\bf 68}, 214509 (2003).



\bibitem{KCh}
A. B. Klimov and S. M. Chumakov, {\it A Group-Theoretical Approach to Quantum Optics}, (Wiley-VCH, 2009).


\bibitem{victoralbert}
V. V. Albert, G. D. Scholes, and P. Brumer, Phys. Rev. A {\bf 84}, 042110 (2011).




\bibitem{rich}
R. Richardson, Phys. Lett. {\bf 3}, 277 (1963).

\bibitem{rich2}
R. Richardson and N. Sherman, Nucl. Phys. {\bf 52}, 221 (1964).

\end{thebibliography}
\end{document}